# A Warm-start QAOA based approach using a swap-based mixer for the TSP: theoretical considerations, implementation and experiments

Eric Bourreau[1], Gérard Fleury[2] and Philippe Lacomme[2]

[1] *Université Montpellier, LIRMM UMR 5506, CC477, 161 rue Ada, 34095 Montpellier Cedex 5 – France*
Eric.Bourreau@lirmm.fr

[2] *Université Clermont Auvergne, UMR 6158 LIMOS, 1 rue de la Chébarde, Aubière, 63178, France*
Gerard.fleury@isima.fr, placomme@isima.fr

## Abstract

This paper investigates quantum heuristics based on Mixer Hamiltonians, which allow the search to be restricted to a specific subspace and enable warm-start strategies for solving the Traveling Salesman Problem (TSP). Approaches involving Mixer Hamiltonians can be integrated into the Quantum Approximate Optimization Algorithm (QAOA), where the Mixer acts as a mapping function that transforms qubit strings into feasible solution sets. We first introduce a swap-based mixer tailored to the TSP, which ensures that only qubit strings representing valid TSP solutions are explored during the QAOA process. Second, we propose a warm-start technique that initializes QAOA with a solution generated by any classical heuristic, thereby promoting faster convergence. These two contributions are combined into a Warm-Start QAOA framework with a Swap-Based Mixer, leveraging both structural and initialization advantages. Experimental results on a custom TSP instance involving five customers demonstrate the effectiveness of this approach, providing, for the first time, a viable integration of warm-start and swap-based mixers for the TSP within a quantum optimization framework.

## 1 Introduction

Quantum annealing has been proposed as a heuristic technique for leveraging quantum mechanical effects to address discrete optimization problems. These problems typically involve optimizing a quadratic cost function subject to a set of linear constraints. A common strategy for handling these constraints is to incorporate them into the cost function as penalty terms, thereby converting the constrained optimization problem into an unconstrained one. However, this method presents certain drawbacks, including increased resource demands, such as higher connectivity and an expanded dynamical range for the parameters defining the problem instance.

Lately it has been introduced the concept of Constrained Quantum Annealing (CQA), which employs designed driver Hamiltonians tailored to specific constraints. These tuned Hamiltonians offer several advantages, including a reduction in the problem's search space and a decrease in the number of interactions required for implementing the annealing algorithm. Central to this approach is the principle that a Hamiltonian commuting with the operator representation of the constraints and initialized within the feasible configuration space will remain within it throughout the evolution. CQA has been successfully applied to the graph coloring problem by leveraging a specific Hamiltonian that ensures each vertex is assigned a unique color. This greatly simplifies the modeling process, as it then only requires an objective function aimed at minimizing the number of conflicts between adjacent vertices and a single constraint.

The earliest contribution comes from (Hadfield, 2018), who proposed a Driver Hamiltonian for the graph coloring problem. The usual formulation involves an objective function composed of two elements: the minimization of the number of colors and a measure of constraint violations penalized by a coefficient $A$. The formulation is based on binary variables $x_{ik}$ where $x_{ik} = 1$ if color $k \in [1, K]$ is assigned to vertex $i$, with $K$ being the number of colors and $n$ the number of vertices (considering $K$ the number of colors, $n$ the number of nodes, $E$ the set of arcs) .

$$\text{Min} \sum_{k=1}^{K} \sum_{(i,j)\in E} x_{ik} . x_{jk} + A . \sum_{i=1}^{n} \left(1 - \sum_{k=1}^{K} x_{ik}\right)^2$$

The classical modelization is based on the mixer

$$H_D = -\sum_{i=1}^{n} \sum_{j=1}^{K} X_{ij}$$

and on $H_P$ with

$$H_P = \sum_{k=1}^{K} \sum_{(i,j)\in E} \frac{1}{2}(Id - Z_{ik}) . \frac{1}{2}(Id - Z_{jk}) + A . \sum_{i=1}^{n} \left(1 - \sum_{k=1}^{K} \frac{1}{2}(Id - Z_{ik})\right)^2$$

The classical $H_D$ only defines a uniform distribution of qubit-strings assigning the minimal expected value to meet the adiabatic constraint. The Driver Hamiltonian proposed by Hadfield consists of superposing all states in which a unique color is assigned to each vertex:

$$H_D = \sum_{i=1}^{n} \sum_{j=1}^{K} \left(X_{ij} X_{ij+1} + Y_{ij} Y_{ij+1}\right)$$

And $H_p$ modelizes the mininimization of the color assignment conflict between vertex connected by one arc.

$$H_p = \sum_{(i,j)\in E} \sum_{k=1}^{K} \frac{1}{4}\left(Id + Z_{ik} + Z_{jk} + Z_{ik} . Z_{jk}\right)$$

Numerical experiments prove the efficiency of such a modelization in (Bourreau et al., 2022). Many research studies have been conducted in this same line of research initiated by Hadfield. Experiments carried out by (Govia et al., 2021) confirm the interest of this type of mixer and the conclusions were confirmed in 2021 by (Shaydulin et al., 2021) and in 2022 by (Golden et al., 2022) et (Zhu et al., 2022). Lately (Tsvelikhovskiy et al., 2024) gives a systematic methodology for constructing QAOA tailored mixer Hamiltonian taking advantages of classical symmetries group inherent in the objective function.

According to (Tsvelikhovskiy et al., 2024) dedicated mixers must favor the exploitation of optimization problem structure to improve speed-ups and should meet several objectives including for example:

**Enforcement of hard constraint**s to enforce qubit-string modelization of solutions only. Significant improvement have been prove using the $XY$-mixer of Hadfield for the graph-coloring;
**Alignment with initial state** i.e. the initial state of QAOA have to be set to the ground state of the mixing Hamiltonian, as required by the adiabatic algorithm to improve QAOA efficiency.
**Exploiting Problem Structure for Speed-Up to** optimal mixers dedicated to optimization problems to increase the computational speed-up.

After (Hadfield, 2018) who has introduced the commutation condition between $H_D$ and the initial state, (Tsvelikhovskiy et al., 2024) introduces an approach to construct a mixer Hamiltonian for QAOA that aligns with the symmetries of the objective function.

This paper focuses on the definition of swap based mixer for the TSP. It is structured as follows: section 2 presents formulations of the main results and introduces properties for the swap based mixer. The subsequent section introduces numerical experiments using Qiskit using a 5-customers instance to provide a verification of the claim properties.

# 2 Proposition

## 2.1 TSP modeling

The Traveling Salesman Problem (TSP) is a vehicle routing problem in which a set of customers must be visited by a vehicle, without necessarily requiring a pickup or delivery at each stop. The objective of this transportation problem is to minimize the total distance traveled by the vehicle. In this type of problem, the only given data is the distance matrix $w_{j,k}$, which represents the distance between two customers $j$ and $k$. Let $T$ be a tour in a TSP problem with $n$ vertices, the tour $T$ is defined by the sequence $T_j$, which specifies the customer located at position $j$ in the tour. The cost of the tour is then given by: $C(T) = \sum_{j=0}^{n} w_{T_j,T_{j+1}}$ .

We consider here the example given by (Papalitsas et al., 2019), which includes 3 clients plus the depot denoted as 0. The distance matrix is the one proposed in Table 1, and it corresponds to the graph in Fig. 1. A tour $T$, which is a solution to the problem, is defined by an ordered sequence of nodes. For example, $=$ [0,2,1,3,0], as shown in the diagram below, represents the tour 0,2,1,3,0. The cost of this tour is given by:

$$C(T) = w_{T_0,T_1} + w_{T_1,T_2} + w_{T_2,T_3} + w_{T_3,T_4}$$

$$C(T) = w_{0,2} + w_{2,1} + w_{1,3} + w_{3,0}$$

$$C(T) = 6.11$$

Note that $T[0] = 0$ and $T[n+1] = 0$, meaning that the tour starts and ends at the depot. This type of modeling is commonly used in the Constraint Programming community, as it efficiently handles the issue of subtours, which is often encountered in classical flow-based models."

**Table 1** Example of distances

| | **0** | **1** | **2** | **3** |
|---|---|---|---|---|
| **0** | 0 | 1 | 1.44 | 2.23 |
| **1** | 1 | 0 | 1 | 1.41 |
| **2** | 1.41 | 1 | 0 | 1 |
| **3** | 2.23 | 1.41 | 1 | 0 |

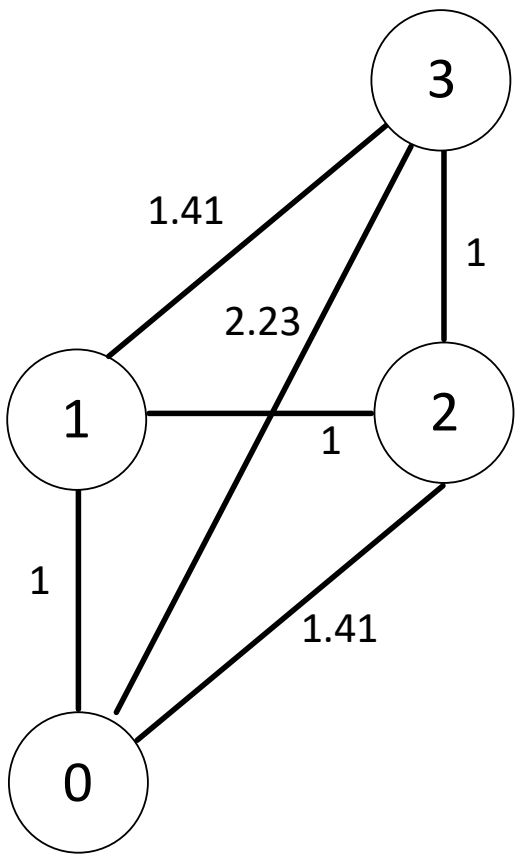

**Fig 1** Graph visualization

The classical modeling uses variables $x_{i,p}$ such that $x_{i,p} = 1$ if customer $i$ is in position $p$ in the tour, and 0 otherwise. The formalization requires four constraints.

Constraint 1. Only One single customer $i$ is assigned to each position $p$.

$$\forall p = 1..n-1, \ \sum_{i=0}^{n-1} \boldsymbol{x_{ip}} = 1$$

Constraint 2. Only One customer $i$ is assigned to one single position $p$.

$$\forall i = 1..n-1, \quad \sum_{p=1}^{n-1} \boldsymbol{x_{ip}} = 1$$

Constraint 3. A trip start at customer 0

$$x_{00} = 1$$

Constraint 4. A trip ends at customer 0

$$x_{0n} = 1$$

The total cost is defined by:

$$C(x) = \sum_{i=0}^{n-1} \sum_{j=0}^{n-1} w_{ij} \, . \sum_{p=1}^{n-2} x_{ip} . x_{j,p+1} + \sum_{i=0}^{n-1} w_{0,i} . x_{i,1} + \sum_{i=0}^{n-1} w_{i,0} . x_{i,n-1}$$

For one instance with $n = 4$ customers, the model introduced at Fig. 2 leading to the following objective function.

$$C(x) = \sum_{i=0}^{3} \sum_{j=0}^{3} w_{ij} \, . \sum_{p=1}^{2} x_{ip} . x_{j,p+1} + \sum_{i=0}^{3} w_{0i} . x_{i,1} + \sum_{i=0}^{3} w_{i,0} . x_{i,3}$$

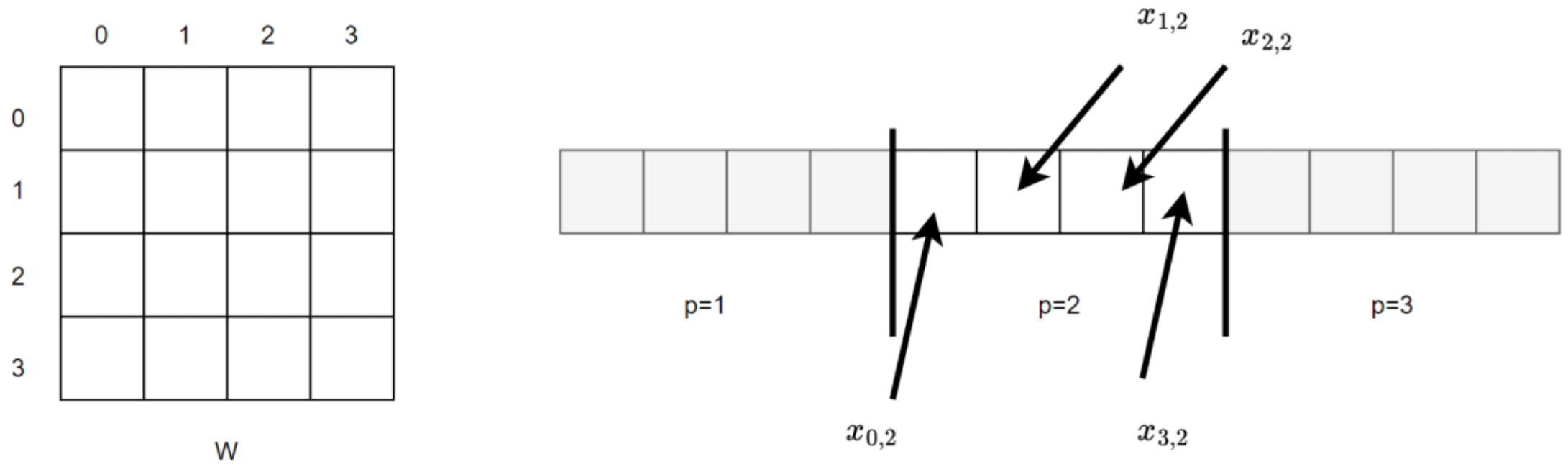

**Fig. 2** Modelisation for a 4 customer-instance

Let us consider the $w_{ij}$:

$$w = \begin{pmatrix} 0 & 1 & 1.44 & 2.23 \\ 1 & 0 & 1 & 1.41 \\ 1.44 & 1 & 0 & 1 \\ 2.23 & 1.41 & 1 & 0 \end{pmatrix}$$

We have

$$\sum_{i=0}^{3} w_{0i}.x_{i,1} = w_{00}.x_{0,1} + w_{01}.x_{1,1} + w_{02}.x_{21} + w_{03}.x_{31}$$

$$\sum_{i=0}^{3} w_{0i}.x_{i,1} = 1.x_{1,1} + 1.44.x_{21} + 2.23.x_{31}$$

and

$$\sum_{i=0}^{3} w_{i,0}.x_{i,3} = w_{0,0}.x_{03} + w_{10}.x_{13} + w_{20}.x_{23} + w_{30}.x_{33}$$

$$\sum_{i=0}^{3} w_{i,0}.x_{i,3} = 1.x_{13} + 1.44.x_{23} + 2.23.x_{33}$$

and

$$\sum_{i=0}^{3}\sum_{j=0}^{3} w_{ij}.\sum_{p=1}^{2} x_{ip}.x_{j,p+1}$$

$$= w_{00}.\left(x_{01}.x_{0,2} + x_{02}.x_{0,3}\right) + w_{01}.\left(x_{01}.x_{1,2} + x_{02}.x_{1,3}\right) + w_{02}.\left(x_{01}.x_{2,2} + x_{02}.x_{2,3}\right) + w_{03}.\left(x_{01}.x_{3,2} + x_{02}.x_{3,3}\right)$$

$$+w_{10}.\left(x_{11}.x_{0,2} + x_{12}.x_{0,3}\right) + w_{11}.\left(x_{11}.x_{1,2} + x_{12}.x_{1,3}\right) + w_{12}.\left(x_{11}.x_{2,2} + x_{12}.x_{2,3}\right) + w_{13}.\left(x_{11}.x_{3,2} + x_{12}.x_{3,3}\right)$$

$$+w_{20}.\left(x_{21}.x_{0,2} + x_{22}.x_{0,3}\right) + w_{21}.\left(x_{21}.x_{1,2} + x_{22}.x_{1,3}\right) + w_{22}.\left(x_{21}.x_{2,2} + x_{22}.x_{2,3}\right) + w_{23}.\left(x_{21}.x_{3,2} + x_{22}.x_{3,3}\right)$$

$$+w_{30}.\left(x_{31}.x_{0,2} + x_{32}.x_{0,3}\right) + w_{31}.\left(x_{31}.x_{1,2} + x_{32}.x_{1,3}\right) + w_{23}.\left(x_{21}.x_{3,2} + x_{22}.x_{3,3}\right) + w_{33}.\left(x_{31}.x_{3,2} + x_{32}.x_{3,3}\right)$$

This defines a complex function with a large number of terms. The unconstraint formalization consists in defined a new objective function that encompasses two extra penalization terms that are supposed to measure the violation of constraint using a weigh $A$ that could be instance dependent. Let us note that constraint 3 and constraint 4 vanished since we have only to modelize the customers from position 2 to position $n-1$.

$$C(x) = \sum_{i=0}^{n-1}\sum_{j=0}^{n-1} w_{ij} \cdot \sum_{p=1}^{n-2} x_{ip}.x_{j,p+1} + \sum_{i=0}^{n-1} w_{0,i}.x_{i,1} + \sum_{i=0}^{n-1} w_{i,0}.x_{i,n-1} + A.\sum_{p=1}^{n-1}\left(1-\sum_{i=1}^{n} \boldsymbol{x_{ip}}\right)^2 + B.\sum_{i=1}^{n}\left(\mathbf{1}-\sum_{p=1}^{n-1} \boldsymbol{x_{ip}}\right)^{\mathbf{2}}$$

Such approach permits to define a quadratic TSP formulation where the major drawback consists in defining a search space that is composed of all the qubit-strings of size $n \times (n-1)$ and where one qubit-string can defined either one TSP solution or one trip where constraint 3 or 4 can be unsatisfied.

## 2.2 Driver Hamiltonian

The principle consists in defining an operator $H_B$ that puts all the solutions of the TSP into superposition, and only the solutions of the TSP, meaning the qubit-strings that satisfy constraints 3 and 4. To define this operator, we need to consider the operator $Swap_{ij}$, which swaps two qubits and that is defined as:

$$Swap_{ij} = CX_{ij}.CX_{ji}.CX_{ij}$$

With the operator $Swap_{ij}$, one could then imagine easily transforming one trip into another in which two customer have been swapped. For example, one could transform the solution $0-2-1-3-0$ from Fig. 3 into a new solution 0-1-2-3-0 (Fig. 4) by simply swapping customer 2 and customer 1

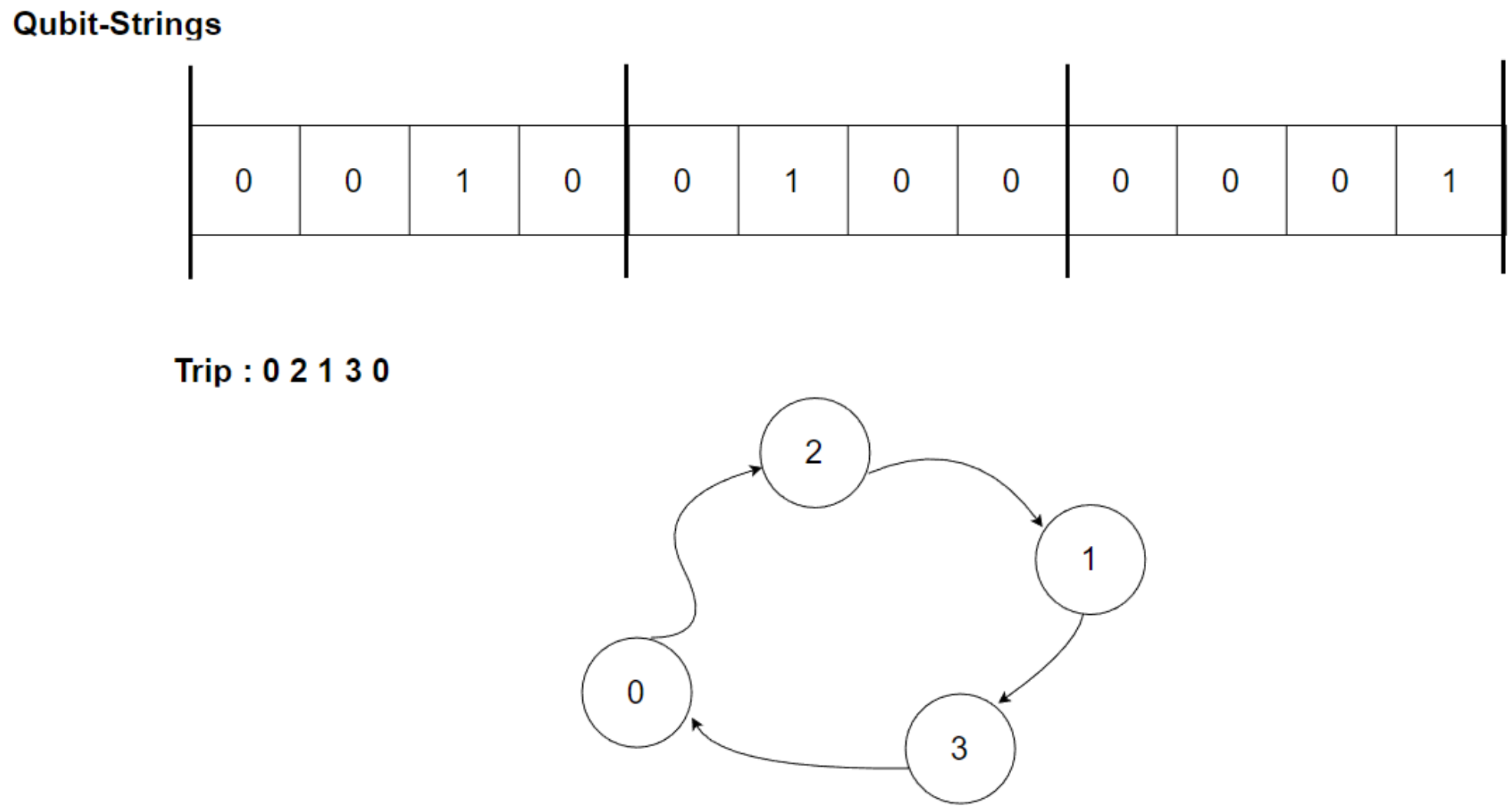

**Fig. 3** Visualization of the solution $0-2-1-3-0$

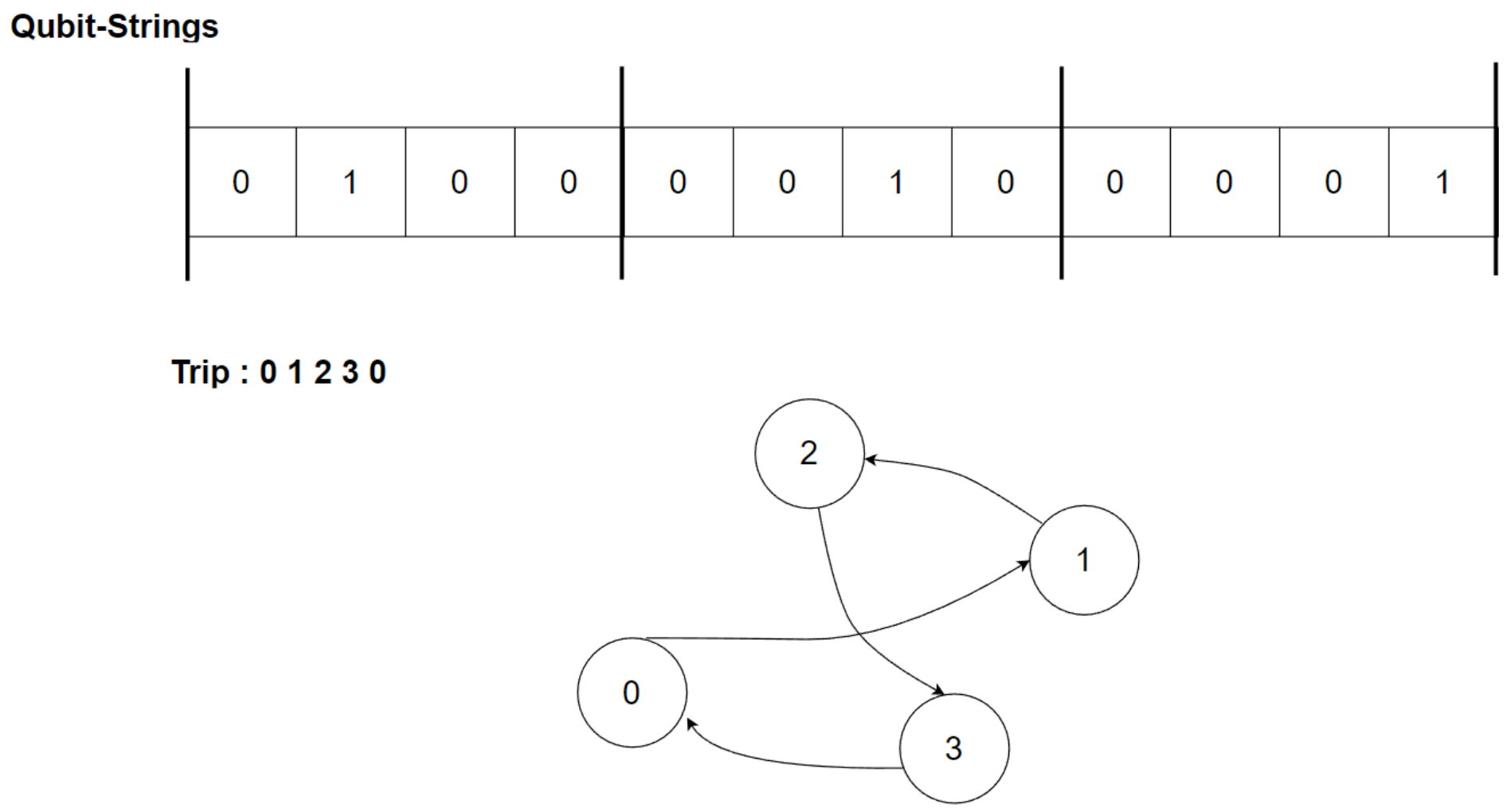

**Fig. 4** Visualization of the solution $0-1-2-3-0$

Thus, swapping the element in the first position of the tour with the second element consists, in a first analysis, of swapping qubits $q_0$ and $q_4$, which correspond to the variables $x_{00}$ and $x_{10}$, then swapping qubits $q_1$ and $q_5$, qubits $q_2$ and $q_6$, and finally swapping qubits $q_3$ et $q_7$, as shown in Fig. 5. We will show that this is not so simple in the following paragraph.

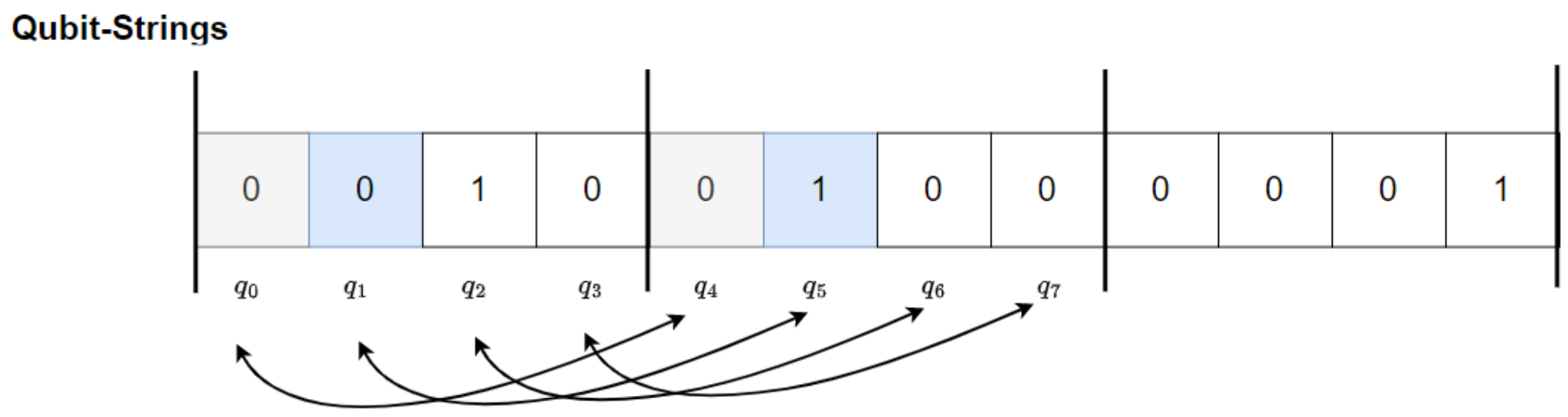

**Fig. 5** Successive application of $Swap_{ij}$ operator

## 2.3 Swap definition

Let us consider a two qubit Hilbert space where $CX_{12}.CX_{21}.CX_{12} = swap_{12}$.
Let

$$|\psi_1\rangle = a.|0\rangle + b.|1\rangle$$
$$|\psi_2\rangle = c.|0\rangle + d.|1\rangle$$

We have:

$$|\psi\rangle = |\psi_1\rangle \otimes |\psi_2\rangle$$
$$|\psi\rangle = (a.|0\rangle + b.|1\rangle) \otimes (c.|0\rangle + d.|1\rangle)$$
$$|\psi\rangle = ac.|00\rangle + ad.|01\rangle + bc.|10\rangle + bd.|11\rangle$$

and

$$|\psi_1\rangle = CX_{12}.|\psi\rangle = ac.|00\rangle + ad.|01\rangle + bc.|11\rangle + bd.|10\rangle$$

and

$$|\psi_2\rangle = CX_{21}.|\psi_1\rangle = ac.|00\rangle + ad.|11\rangle + bc.|01\rangle + bd.|10\rangle$$

and

$$|\psi_3\rangle = CX_{12}.|\psi_2\rangle = ac.|00\rangle + ad.|10\rangle + bc.|01\rangle + bd.|11\rangle$$

and to conclude:

$$|\psi_3\rangle = ac.|00\rangle + ad.|10\rangle + bc.|01\rangle + bd.|11\rangle$$
$$|\psi_3\rangle = c.|0\rangle \otimes (a.|0\rangle + b.|1\rangle) + d.|1\rangle \otimes (a.|0\rangle + b.|1\rangle)$$
$$|\psi\rangle = (c.|0\rangle + d.|1\rangle) \otimes (a.|0\rangle + b.|1\rangle)$$
$$|\psi\rangle = |\psi_2\rangle \otimes |\psi_1\rangle$$

□

In this very specific case it is possible to make a matrix computation for:

$$Swap_{12} = CX_{12}.CX_{21}.CX_{12}$$

with

$$CX_{12} = \begin{pmatrix} 1 & & & \\ & 1 & & \\ & & & 1 \\ & & 1 & \end{pmatrix}$$

and

$$CX_{21} = \begin{pmatrix} 1 & & & \\ & & & 1 \\ & & 1 & \\ & 1 & & \end{pmatrix}$$

So we have:

$$Swap_{12} = \begin{pmatrix} 1 & & & \\ & 1 & & \\ & & & 1 \\ & & 1 & \end{pmatrix} . \begin{pmatrix} \mathbf{1} & & & \\ & & & \mathbf{1} \\ & & \mathbf{1} & \\ & \mathbf{1} & & \end{pmatrix} . \begin{pmatrix} 1 & & & \\ & 1 & & \\ & & & 1 \\ & & 1 & \end{pmatrix}$$

$$Swap_{12} = \begin{pmatrix} 1 & & & \\ & 1 & & \\ & & & 1 \\ & & 1 & \end{pmatrix} . \begin{pmatrix} 1 & & & \\ & & 1 & \\ & & & 1 \\ & 1 & & \end{pmatrix}$$

$$Swap_{12} = \begin{pmatrix} 1 & & & \\ & & 1 & \\ & 1 & & \\ & & & 1 \end{pmatrix}$$

Here we have the matrix expression of the $Swap_{12}$ considering a two qubit Hilbert space.
Let us note that:

$$\left(Swap_{12}\right)^2 = \begin{pmatrix} 1 & & & \\ & & 1 & \\ & 1 & & \\ & & & 1 \end{pmatrix} . \begin{pmatrix} 1 & & & \\ & & 1 & \\ & 1 & & \\ & & & 1 \end{pmatrix} = \begin{pmatrix} 1 & & & \\ & 1 & & \\ & & 1 & \\ & & & 1 \end{pmatrix} = Id$$

Because $\left(Swap_{12}\right)^2 = Id$, we have:

$$e^{-i.t.Swap_{12}} = \cos t\,.Id - i.\sin t\,.Swap_{12}$$

and

$$e^{-i.t.Swap_{12}} = \begin{pmatrix} \cos t - i.\sin t & & & \\ & \cos t & -i.\sin t & \\ & -i.\sin t & \cos t & \\ & & & \cos t - i.\sin t \end{pmatrix}$$

Whatever the number of qubit we have:

$$e^{-i.t.Swap_{kp}} = \left[\cos t.Id - i.\sin t\,.\left(Swap_{kp}\right)\right]$$

Let us consider for example a trip composed of 4 positions

We have:

$$e^{-i.t.Swap_{12}}.\,|abcd\rangle = [\cos t.\,Id - i.\sin t\,.\,(Swap_{12})]\;|abcd\rangle$$

$$e^{-i.t.Swap_{12}}.\,|abcd\rangle = \cos t.\,|abcd\rangle - i.\sin t\,.\,|bacd\rangle$$

One can conclude that the operator $e^{-i.t.Swap_{12}}$ applied to $|abcd\rangle$ defines a list of two quantum states including $|abcd\rangle$ and $|bacd\rangle$.

And next:

$$e^{-i.t.Swap_{23}}.\,e^{-i.t.Swap_{12}}.\,|abcd\rangle = e^{-i.t.Swap_{23}}(\cos t.\,|abcd\rangle - i.\sin t\,.\,|bacd\rangle\,)$$

$$e^{-i.t.Swap_{23}}.\,e^{-i.t.Swap_{12}}.\,|abcd\rangle = [\cos t.\,Id - i.\sin t\,.\,(Swap_{23})].\,(\cos t.\,|abcd\rangle - i.\sin t\,.\,|bacd\rangle\,)$$

$$e^{-i.t.Swap_{23}}.\,e^{-i.t.Swap_{12}}.\,|abcd\rangle = \cos t.\cos t.\,|abcd\rangle - i.\sin t\,.\cos t.\,|bacd\rangle - i.\cos t.\sin t\,|acbd\rangle - \sin t\,.\sin t\;\;|bcad\rangle$$

$$e^{-i.t.Swap_{23}}.\,e^{-i.t.Swap_{12}}.\,|abcd\rangle = \cos^2 t.\,|abcd\rangle - i.\sin t\,.\cos t.\,|bacd\rangle - i.\cos t.\sin t\,|acbd\rangle - \sin^2 t\;\;|bcad\rangle$$

and

$$e^{-i.t.Swap_{34}}.\,e^{-i.t.Swap_{23}}.\,e^{-i.t.Swap_{12}}.\,|abcd\rangle = [\cos t.\,Id - i.\sin t\,.\,(Swap_{34})](\cos^2 t.\,|abcd\rangle - i.\sin t\,.\cos t.\,|bacd\rangle - i.\cos t.\sin t\,|acbd\rangle - \sin^2 t\;\;|bcad\rangle)$$

$$e^{-i.t.Swap_{34}}.\,e^{-i.t.Swap_{23}}.\,e^{-i.t.Swap_{12}}.\,|abcd\rangle = \cos^3 t.\,|abcd\rangle - i.\sin t\,.\cos^2 t.\,|bacd\rangle - i.\cos^2 t.\sin t\,|acbd\rangle - \cos t.\sin^2 t\;\;|bcad\rangle - i.\cos^2 t.\sin t\,.\,|abdc\rangle - \sin^2 t\,.\cos t.\,|badc\rangle - \cos t.\sin^2 t\,|acdb\rangle + i\sin^3 t\;\;|bcda\rangle$$

and

$$e^{-i.t.Swap_{41}}.\,e^{-i.t.Swap_{34}}.\,e^{-i.t.Swap_{23}}.\,e^{-i.t.Swap_{12}}.\,|abcd\rangle = [\cos t.\,Id - i.\sin t\,.\,(Swap_{41})](\cos^3 t.\,|abcd\rangle - i.\sin t\,.\cos^2 t.\,|bacd\rangle - i.\cos^2 t.\sin t\,|acbd\rangle - \cos t.\sin^2 t\;\;|bcad\rangle - i.\cos^2 t.\sin t\,.\,|abdc\rangle - \sin^2 t\,.\cos t.\,|badc\rangle - \cos t.\sin^2 t\,|acdb\rangle + i\sin^3 t\;\;|bcda\rangle)$$

So we can conclude that:

$$e^{-i.t.Swap_{41}}.\,e^{-i.t.Swap_{34}}.\,e^{-i.t.Swap_{23}}.\,e^{-i.t.Swap_{12}}.\,|abcd\rangle = cos^4\,t.\,|abcd\rangle - i.\,sin\,t\,.\,cos^3\,t.\,|bacd\rangle - i.\,cos^3\,t.\,sin\,t\,|acbd\rangle - cos^2\,t.\,sin^2\,t\;\;|bcad\rangle - i.\,cos^3\,t.\,sin\,t\,.\,|abdc\rangle - sin^2\,t\,.\,cos^2\,t.\,|badc\rangle - cos^2\,t.\,sin^2\,t\,|acdb\rangle + i\,sin^3\,t\,.\,cos\,t\;\;|bcda\rangle - i.\,cos^3\,t.\,sin\,t\,.\,|dbca\rangle - \sin^2 t\,.\cos^2 t.\,|dacb\rangle - \cos^2 t.\sin^2 t\,|dcba\rangle + i.\cos t.\sin^3 t\;\;|dcab\rangle - \cos^2 t.\sin^2 t\,.\,|cbda\rangle + i.\sin^3 t\,.\cos t.\,|cadb\rangle + i.\cos t.\sin^3 t\,|bcda\rangle + \sin^4 t\;\;|acdb\rangle$$

## Conclusion

The Hamiltonian $H_B = Swap_{41} + Swap_{34} + Swap_{23} + Swap_{12}$, permits to define a quantum state composed of 16 permutations.

**Remark**

$$Swap_{ij} = \frac{1}{2}.Id + \frac{1}{2}X_iX_j + \frac{1}{2}.Y_i.Y_j + \frac{1}{2}.Z_i.Z_j$$

In the special case of two-qubits circuit we have:

$$X_1 = \begin{pmatrix} & & 1 & \\ & & & 1 \\ 1 & & & \\ & 1 & & \end{pmatrix} \text{ and } X_2 = \begin{pmatrix} & 1 & & \\ 1 & & & \\ & & & 1 \\ & & 1 & \end{pmatrix}$$

$$Y_1 = \begin{pmatrix} & & -i & \\ & & & -i \\ i & & & \\ & i & & \end{pmatrix} \text{ and } Y_2 = \begin{pmatrix} & -i & & \\ i & & & \\ & & & -i \\ & & i & \end{pmatrix}$$

$$Z_1 = \begin{pmatrix} 1 & & & \\ & 1 & & \\ & & -1 & \\ & & & -1 \end{pmatrix} \text{ and } Z_2 = \begin{pmatrix} 1 & & & \\ & -1 & & \\ & & 1 & \\ & & & -1 \end{pmatrix}$$

And

$$X_1X_2 = \begin{pmatrix} & & & 1 \\ & & 1 & \\ & 1 & & \\ 1 & & & \end{pmatrix}$$

$$Y_1Y_2 = \begin{pmatrix} & & & -1 \\ & & 1 & \\ & 1 & & \\ -1 & & & \end{pmatrix}$$

$$Z_1Z_2 = \begin{pmatrix} 1 & & & \\ & -1 & & \\ & & -1 & \\ & & & 1 \end{pmatrix}$$

And by consequence

$$Swap_{12} = \frac{1}{2}.Id + \frac{1}{2}X_1X_2 + \frac{1}{2}.Y_1.Y_2 + \frac{1}{2}.Z_1.Z_2$$

$$Swap_{12} = \frac{1}{2}.\begin{pmatrix} 1 & & & \\ & 1 & & \\ & & 1 & \\ & & & 1 \end{pmatrix} + \frac{1}{2}\begin{pmatrix} & & & 1 \\ & & 1 & \\ & 1 & & \\ 1 & & & \end{pmatrix} + \frac{1}{2}.\begin{pmatrix} & & & -1 \\ & & 1 & \\ & 1 & & \\ -1 & & & \end{pmatrix}$$
$$+ \frac{1}{2}.\begin{pmatrix} 1 & & & \\ & -1 & & \\ & & -1 & \\ & & & 1 \end{pmatrix}$$

$$Swap_{12} = \frac{1}{2}.\begin{pmatrix} 2 & & & \\ & & 2 & \\ & 2 & & \\ & & & 2 \end{pmatrix}$$

## 2.4 Binary-Swap application to TSP

In a first analysis, a Swap operator applied to the TSP, using the example presented in Fig. 3 and 4, to swap the elements at positions 1 and 2, could be defined as:

$$H = Swap_{04} + Swap_{15} + Swap_{26} + Swap_{37}$$

This corresponds to:

$$e^{-i.t.H} \simeq e^{-i.t.Swap_{04}}.e^{-i.t.Swap_{14}}.e^{-i.t.Swap_{26}}.e^{-i.t.Swap_{37}}$$

applied to $|00100100\rangle$, following the same example as in Fig. 4. It is then possible to perform the calculations since: $e^{-i.t.Swap} = [\cos t.Id - i.\sin t.(Swap)]$

Thus, without representing the associated probabilities:

$$e^{-i.t.H}.|00100100\rangle \simeq e^{-i.t.Swap_{04}}.e^{-i.t.Swap_{14}}.e^{-i.t.Swap_{26}}.e^{-i.t.Swap_{37}}.|001000100\rangle$$

$$e^{-i.t.H}.|00100100\rangle \simeq e^{-i.t.Swap_{04}}.e^{-i.t.Swap_{14}}.e^{-i.t.Swap_{26}}.|00100100\rangle$$

$$e^{-i.t.H}.|00100100\rangle \simeq e^{-i.t.Swap_{04}}.e^{-i.t.Swap_{14}}.(|00100100\rangle + |00000110\rangle)$$

This results in a final state that does not correspond to the permutation of the elements in positions 0-3 with those in positions 4-7. To perform a Swap of 4 qubits with 4 other qubits, it is necessary to define a global operator that executes the operation rather than decomposing this swap into 4 elementary swaps. Thus, a swap for qudits must be defined.

## 2.5 Modelizations

Considering one operator $U$ such that $U^2 = Id$ it is possible to compute $e^{-i.t.U}$ considering the Algorithm 1.

**Algorithm 1. Definition of $e^{-i.t.U}$**

Let $U$ such that $U^2 = Id$, meaning that $e^{-i.t.U} = \cos t.Id - i.\sin t.U$.

Assume that $U$ is a n-dimensionnal operator applied to $|\psi_1, \ldots, \psi_{n-1}, \psi_n\rangle$.

Let $|\psi_0\rangle$ one auxiliary qubit

First defines: $|\psi_0\rangle = \cos t.|0\rangle - i.\sin t.|1\rangle$
So $CU_0.|\psi_0, \psi_1, \ldots, \psi_{n-1}, \psi_n\rangle = \cos t.|0, \psi_1, \ldots, \psi_{n-1}, \psi_n\rangle - i.\sin t.|1\rangle \otimes U.|\psi_1, \ldots, \psi_{n-1}, \psi_n\rangle$

By consequence the qubits 1 to $n$ defined now:

$$(\cos t.Id - i.\sin t.U).|\psi_1, \ldots, \psi_{n-1}, \psi_n\rangle = e^{-i.t.U}.|\psi_1, \ldots, \psi_{n-1}, \psi_n\rangle$$

□

**Remark**

Since $RX(t) = \begin{pmatrix} \cos\frac{\theta}{2} & -i.\sin\frac{\theta}{2} \\ -i.\sin\frac{\theta}{2} & \cos\frac{\theta}{2} \end{pmatrix}$, we have $|\psi_0\rangle = RX(t).|0\rangle$

**Algorithm 2. Definition of** $C\left(e^{-i.t.U}\right)_{i,j}$

Let $|\psi_0\rangle$ a control-qubit

Assume that $U$ is a n-dimensionnal operator applied to $|\psi_1, \dots, \psi_{n-1}, \psi_n\rangle$.

Using algorithm 1, one can defined:

$$(\cos t\,.Id - i.\sin t\,.U).|\psi_1, \dots, \psi_{n-1}, \psi_n\rangle = e^{-i.t.U}.|\psi_1, \dots, \psi_{n-1}, \psi_n\rangle$$

We have (Fleury and Lacomme, 2025):

$$e^{-i.t.U} = <> RK(\theta) <> \text{ with } RK(\theta) = RX(\theta) \text{ or } RK(\theta) = RZ(\theta)$$

where $<>$ is a list of gate that depends on $U$.

So to define $Ce^{-i.t.U}$ we just have to replace $RK(\theta)$ by $CRK(\theta)$ using the control-qubit $|\psi_0\rangle$ :

$$C\left(e^{-i.t.U}\right)_{0,1} = <> CRK_{0,1}(\theta) <>$$

□

**Algorithm 3. Definition of** $\boldsymbol{U}$

We have $e^{-i.t.U} = \cos t\,.Id - i.\sin t\,.U$

And using $t = \frac{\pi}{2}$, we have: $e^{-i.\frac{\pi}{2}.U} = -i.U$

because $P(\theta) = \begin{pmatrix} 1 & \\ & e^{i.\theta} \end{pmatrix}$, we have $P\left(\frac{\pi}{2}\right) = \begin{pmatrix} 1 & \\ & i \end{pmatrix}$

and $P\left(\frac{\pi}{2}\right).\begin{pmatrix} a \\ b \end{pmatrix} = \begin{pmatrix} a \\ i.b \end{pmatrix}$

So $P\left(\frac{\pi}{2}\right).X.P\left(\frac{\pi}{2}\right).X.\begin{pmatrix} a \\ b \end{pmatrix} = i.\begin{pmatrix} a \\ b \end{pmatrix}$

and

$$P\left(\frac{\pi}{2}\right).X.P\left(\frac{\pi}{2}\right).X.e^{-i.\frac{\pi}{2}.U} = U$$

$$i.e^{-i.\frac{\pi}{2}.U} = U$$

□

**Algorithm 4. Definition of $CU$**

$$CU_{0,1} = CP_{0,1}\left(\frac{\pi}{2}\right).CX_{0,1}.CP_{0,1}\left(\frac{\pi}{2}\right).CX_{0,1}.C\left(e^{-i.t.U}\right)_{0,1}$$

▯

## 2.6 Decomposition of operator into Pauli gates

Let us consider the linearly independent set of operator $(Id, X, Y, Z)$. The vector space generated consists of $2 \times 2$ matrices of the form:

$$A = a.Id + b.X + c.Y + d.Z$$

where $a, b, c, d$ are complex number

We have

$$A^{\dagger} = \bar{a}.Id + \bar{b}.X + \bar{c}.Y^{\dagger} + \bar{d}.Z$$

So, $A = A^{\dagger}$ if and only if:

$$\begin{cases} a = \bar{a} \\ b = \bar{b} \\ c.Y = \bar{c}.Y^{\dagger} \\ d = \bar{d} \end{cases}$$

So $a, b, d$ are real numbers and

$$c.\begin{pmatrix} & -i \\ i & \end{pmatrix} = \bar{c}.\begin{pmatrix} & -i \\ i & \end{pmatrix}$$

So $a, b, d$ and $c$ must be real numbers to satisfy the equations.

Assuming that $a, b, d$ and $c$ are real numbers and determine under what condition $A$ is unitary.

$$A^2 = Id \iff (a.Id + b.X + c.Y + d.Z)^2 = Id$$

Or

$$(a.Id + b.X + c.Y + d.Z)^2 = (a^2 + b^2 + c^2 + d^2).Id + 2.a.b.X + 2.a.c.Y + 2.ad.Z$$

$$+b.c.(X.Y + Y.X) + b.d.(X.Z + Z.X) + c.d.(Y.Z + Z.Y)$$

Because

$$X.Y + Y.X = X.Z + Z.X = Z.Y + Y.Z = 0$$

We have

$$(a.Id + b.X + c.Y + d.Z)^2 = (a^2 + b^2 + c^2 + d^2).Id + 2.a.b.X + 2.a.c.Y + 2.ad.Z$$

So $A$ is unitary if and only if:

$$\begin{cases} a^2 + b^2 + c^2 + d^2 = 1 \\ a.b = 0 \\ a.c = 0 \\ a.d = 0 \end{cases}$$

Two situations hold:

$$\begin{cases} b^2 + c^2 + d^2 = 1 \\ a = 0 \end{cases} \text{ or } \begin{cases} a = \pm 1 \\ b = c = d = 0 \end{cases}$$

So we have

$$A = b.X + c.Y + d.Z \text{ with } b^2 + c^2 + d^2 = 1$$

either

$$A = \pm Id$$

Let us consider the external product between two matrices $A$ and $B$:

$$\langle A|B\rangle = \sum_{i,j} (A^\dagger)_{ij}.B_{ij}$$

For any real value $k$ we have:

$$\langle k.A|B\rangle = \sum_{i,j} (k.A^\dagger)_{ij}.B_{ij} = k.\sum_{i,j} (A^\dagger)_{ij}.B_{ij} = \sum_{i,j} (A^\dagger)_{ij}.\left(k.B_{ij}\right) = \langle A|k.B\rangle = k.\langle A|B\rangle$$

So we have:

$$\langle k.A|k.A\rangle = k^2.\langle A|A\rangle$$

*i.e.*

$$\|k.A\| = |k|.\|A\|$$

We have $A \otimes C$:

$$A \otimes C = \begin{pmatrix} & a_{ij}.C & \\ & & \end{pmatrix}$$

So

$$\langle A \otimes C|B \otimes D\rangle = \sum_{i,j} (A^\dagger \otimes C^\dagger)_{ij}.(B \otimes D)_{ij} = \sum_{i,j} \overline{a_{ji}} b_{ij}.\langle C|D\rangle = \langle A|B\rangle.\langle C|D\rangle$$

Example

$$A = \begin{pmatrix} a_{11} & a_{12} \\ a_{21} & a_{22} \end{pmatrix}, B = \begin{pmatrix} b_{11} & b_{12} \\ b_{21} & b_{22} \end{pmatrix}, C = \begin{pmatrix} c_{11} & c_{12} \\ c_{21} & c_{22} \end{pmatrix}, D = \begin{pmatrix} d_{11} & d_{12} \\ d_{21} & d_{22} \end{pmatrix},$$

We have:

$$\langle A|B\rangle.\langle C|D\rangle = (a_{11}.b_{11} + a_{12}.b_{12} + a_{21}.b_{21} + a_{22}.b_{22}).(c_{11}.d_{11} + c_{12}.d_{12} + c_{21}.d_{21} + c_{22}.d_{22})$$

$$\langle A|B\rangle.\langle C|D\rangle = a_{11}.b_{11}.c_{11}.d_{11} + a_{11}.b_{11}.c_{12}.d_{12} + a_{11}.b_{11}.c_{21}.d_{21} + a_{11}.b_{11}.c_{22}.d_{22}$$

$$+a_{12}.b_{12}.c_{11}.d_{11} + a_{12}.b_{12}.c_{12}.d_{12} + a_{12}.b_{12}.c_{21}.d_{21} + a_{12}.b_{12}.c_{22}.d_{22}$$

$$+a_{21}.b_{21}.c_{11}.d_{11} + a_{21}.b_{21}.c_{12}.d_{12} + a_{21}.b_{21}.c_{21}.d_{21} + a_{21}.b_{21}.c_{22}.d_{22}$$

$$+a_{22}.b_{22}.c_{11}.d_{11} + a_{22}.b_{22}.c_{12}.d_{12} + a_{22}.b_{22}.c_{21}.d_{21} + a_{22}.b_{22}.c_{22}.d_{22}$$

and

$$A \otimes C = \begin{pmatrix} a_{11}.C & a_{12}.C \\ a_{21}.C & a_{22}.C \end{pmatrix} = \begin{pmatrix} a_{11}.c_{11} & a_{11}.c_{12} & a_{12}.c_{11} & a_{12}.c_{12} \\ a_{11}.c_{21} & a_{11}.c_{22} & a_{12}.c_{21} & a_{12}.c_{22} \\ a_{21}.c_{11} & a_{21}.c_{12} & a_{22}.c_{11} & a_{22}.c_{12} \\ a_{21}.c_{21} & a_{21}.c_{22} & a_{22}.c_{21} & a_{22}.c_{22} \end{pmatrix}$$

and

$$B \otimes D = \begin{pmatrix} b_{11}.D & b_{12}.D \\ b_{21}.D & b_{22}.D \end{pmatrix} = \begin{pmatrix} b_{11}.d_{11} & b_{11}.d_{12} & b_{12}.d_{11} & b_{12}.d_{12} \\ b_{11}.d_{21} & b_{11}.d_{22} & b_{12}.d_{21} & b_{12}.d_{22} \\ b_{21}.d_{11} & b_{21}.d_{12} & b_{22}.d_{11} & b_{22}.d_{12} \\ b_{21}.d_{21} & b_{21}.d_{22} & b_{22}.d_{21} & b_{22}.d_{22} \end{pmatrix}$$

and

$$\begin{aligned} \langle A \otimes C | B \otimes D \rangle = {} & a_{11}.b_{11}.c_{11}.d_{11} + a_{11}.c_{12}.b_{11}.d_{12} + a_{11}.c_{21}.b_{11}.d_{21} + a_{11}.c_{22}.b_{11}.d_{22} \\ & + a_{12}.c_{11}.b_{12}.d_{11} + a_{12}.c_{12}.b_{12}.d_{12} + a_{12}.c_{21}.b_{12}.d_{21} + a_{12}.c_{22}.b_{12}.d_{22} \\ & + a_{21}.c_{11}.b_{21}.d_{11} + a_{21}.c_{12}.b_{21}.d_{12} + a_{21}.c_{21}.b_{21}.d_{21} + a_{21}.c_{22}.b_{21}.d_{22} \\ & + a_{22}.c_{11}.b_{22}.d_{11} + a_{22}.c_{12}.b_{22}.d_{12} + a_{22}.c_{21}.b_{22}.d_{21} + a_{22}.c_{22}.b_{22}.d_{22} \end{aligned}$$

And we conclude that:

$$\langle A \otimes C | B \otimes D \rangle = \langle A|B \rangle . \langle C|D \rangle$$

Let us note we have,

$$\langle Id|Id \rangle = \langle X|X \rangle = \langle Y|Y \rangle = \langle Z|Z \rangle = 2$$

and

$$\langle Id|X \rangle = \langle Id|Y \rangle = \langle Id|Z \rangle = \langle X|Y \rangle = \langle X|Z \rangle = \langle Y|Z \rangle = 0$$

So the family

$$\left( \frac{1}{\sqrt{2}}.Id, \frac{1}{\sqrt{2}}.X, \frac{1}{\sqrt{2}}.Y, \frac{1}{\sqrt{2}}.Z \right)$$

is orthonormal.

If we consider the set of operators for a two-qubits circuit, we have:

$$(Id \otimes Id, Id \otimes X, Id \otimes Y, Id \otimes Z, X \otimes Id, X \otimes X, X \otimes Y, X \otimes Z, Y \otimes Id, Y \otimes X, Y \otimes Y, Y \otimes Z, Z \otimes Id, Z \otimes X, Z \otimes Y, Z \otimes Z)$$

Each operator $E$ in this set, $\|E\| = 2$ and if we consider $E$ and $F$ two different operator we have $\langle E|F \rangle = 0$

So the set of operators

$$\left( \frac{1}{2}.Id \otimes Id, \frac{1}{2}.Id \otimes X, \frac{1}{2}.Id \otimes Y, \frac{1}{2}.Id \otimes Z, \frac{1}{2}.X \otimes Id, \frac{1}{2}.X \otimes X, \frac{1}{2}.X \otimes Y, \frac{1}{2}.X \otimes Z, \frac{1}{2}.Y \otimes Id, \frac{1}{2}.Y \otimes X, \frac{1}{2}.Y \otimes Y, \frac{1}{2}.Y \otimes Z, \frac{1}{2}.Z \otimes Id, \frac{1}{2}.Z \otimes X, \frac{1}{2}.Z \otimes Y, \frac{1}{2}.Z \otimes Z \right)$$

Is orthonormal.

If we consider one base orthonormal basis $\left(e_j\right)_{1\leq j\leq n}$, then each operator $u$ can be written:

$$u = \sum_{j=1}^{n} \langle u_j | e_j \rangle . e_j$$

**Example**

Let us consider $H = \frac{1}{\sqrt{2}}.\begin{pmatrix} 1 & 1 \\ 1 & -1 \end{pmatrix}$. We have:

$$\left\langle H \middle| \frac{1}{\sqrt{2}}.Id \right\rangle = 0$$

$$\left\langle H \middle| \frac{1}{\sqrt{2}}.X \right\rangle = 1$$

$$\left\langle H \middle| \frac{1}{\sqrt{2}}.Y \right\rangle = 0$$

$$\left\langle H \middle| \frac{1}{\sqrt{2}}.Z \right\rangle = 1$$

And we have the decomposition of $H$ in the basis: $H = \frac{1}{\sqrt{2}}.X + \frac{1}{\sqrt{2}}.Z$

**Example**

Let us consider

$$A = \begin{pmatrix} 1 & & & \\ & 1 & & \\ & & & 1 \\ & & 1 & \end{pmatrix}$$

We have:

$$\left\langle A \middle| \frac{1}{2}.Id \otimes Id \right\rangle = 1$$
$$\left\langle A \middle| \frac{1}{2}.Id \otimes X \right\rangle = 1$$
$$\left\langle A \middle| \frac{1}{2}.Id \otimes Y \right\rangle = 0$$
$$\left\langle A \middle| \frac{1}{2}.Id \otimes Z \right\rangle = 0$$
$$\left\langle A \middle| \frac{1}{2}.X \otimes Id \right\rangle = 0$$
$$\left\langle A \middle| \frac{1}{2}.X \otimes X \right\rangle = 0$$
$$\left\langle A \middle| \frac{1}{2}.X \otimes Y \right\rangle = 0$$
$$\left\langle A \middle| \frac{1}{2}.X \otimes Z \right\rangle = 0$$

$$\left\langle A \middle| \frac{1}{2}.Y \otimes Id \right\rangle = 0$$
$$\left\langle A \middle| \frac{1}{2}.Y \otimes X \right\rangle = 0$$
$$\left\langle A \middle| \frac{1}{2}.Y \otimes Y \right\rangle = 0$$
$$\left\langle A \middle| \frac{1}{2}.Y \otimes Z \right\rangle = 0$$
$$\left\langle A \middle| \frac{1}{2}.Z \otimes Id \right\rangle = 1$$
$$\left\langle A \middle| \frac{1}{2}.Z \otimes X \right\rangle = -1$$
$$\left\langle A \middle| \frac{1}{2}.Z \otimes Y \right\rangle = 0$$
$$\left\langle A \middle| \frac{1}{2}.Z \otimes Z \right\rangle = 0$$

And we conclude that

$$A = \begin{pmatrix} 1 & & & \\ & 1 & & \\ & & & 1 \\ & & 1 & \end{pmatrix}$$

can be decomposed into the basis operators:

$$A = \frac{1}{2}.(Id \otimes Id + Id \otimes X + Z \otimes Id - Z \otimes X)$$

# 3 Warm Start definition

## 3.1 Mixer state definition

The distribution defines by the mixer depends on the Swap operator and on $|\psi_i\rangle$ the initial state that has been choose to generate the distribution.

**Definition of $|\psi_0\rangle$**

The distribution defined by the mixer is

$$|\psi_0\rangle = \prod_{k=1}^{n} e^{-i.\alpha_k.swap_{k,k+1}} . |\psi_b\rangle$$

with $k+1 = (k+1)\ mod\ n$ and $|\psi_b\rangle$ the initial state

▯

For example, let us consider $n = 4$ and $|\psi_b\rangle = |2134\rangle$.

We have

$$|\psi_0\rangle = \prod_{k=1}^{n} e^{-i.\alpha_k.swap_{k,k+1}} . |2134\rangle$$

The successive application of exponential leads first to

$$|\psi_0\rangle = \left(\prod_{k=2}^{n} e^{-i.\alpha_k.swap_{k,k+1}} . |2134\rangle\right).(\cos\alpha_1\ |1234\rangle - i\sin\alpha_1 |2134\rangle)$$

And to conclude:

$$\begin{aligned} |\psi_0\rangle = {} & \cos\alpha_4\cos\alpha_3\cos\alpha_2\cos\alpha_1\ |1234\rangle - i\cos\alpha_4\cos\alpha_3\cos\alpha_2\sin\alpha_1 |2134\rangle \\ & - i.\cos\alpha_4\cos\alpha_3\sin\alpha_2\cos\alpha_1\ |1324\rangle - \cos\alpha_4\cos\alpha_3\sin\alpha_2\sin\alpha_1 |2314\rangle \\ & - i.\cos\alpha_4\sin\alpha_3\cos\alpha_2\cos\alpha_1\ |1243\rangle - \cos\alpha_4\sin\alpha_3\cos\alpha_2\sin\alpha_1 |2143\rangle \\ & - \cos\alpha_4\sin\alpha_3\sin\alpha_2\cos\alpha_1\ |1342\rangle + i.\cos\alpha_4\sin\alpha_3\sin\alpha_2\sin\alpha_1 |2341\rangle \\ & - i.\sin\alpha_4\,(\cos\alpha_3\cos\alpha_2\cos\alpha_1\ |4231\rangle - i\cos\alpha_3\cos\alpha_2\sin\alpha_1 |4132\rangle \\ & - i.\cos\alpha_3\sin\alpha_2\cos\alpha_1\ |4321\rangle - \cos\alpha_3\sin\alpha_2\sin\alpha_1 |4312\rangle \\ & - i.\sin\alpha_3\cos\alpha_2\cos\alpha_1\ |3241\rangle - \sin\alpha_3\cos\alpha_2\sin\alpha_1 |3142\rangle \\ & - \sin\alpha_3\sin\alpha_2\cos\alpha_1\ |2341\rangle + i.\sin\alpha_3\sin\alpha_2\sin\alpha_1 |1342\rangle) \end{aligned}$$

## 3.2 Non quantum methods for TSP

The Traveling Salesman Problem (TSP) has been the subject of extensive research within the scientific community, and the sheer volume of contributions makes it difficult to classify them comprehensively. In this work, we arbitrarily distinguish between heuristic methods and iterative metaheuristic approaches. The **Nearest Neighbour Heuristic** begins by selecting an initial node at random, followed by the iterative selection of the closest unvisited node to the most recently visited one. This process is repeated until the tour returns to the initial node, thereby completing a cycle. In contrast, the **Insertion Heuristic** initiates the tour construction with a partial route comprising a single node and progressively incorporates additional nodes. The cost of inserting a node $k$ between two consecutive nodes $i$ and $j$ in the tour is computed as $c_{ikj} = d_{ik} + d_{kj} - d_{ij}$ , where $d$ denotes the distance metric. Several variants of insertion strategies exist, including **Random Insertion, Nearest Insertion, and Farthest Insertion heuristics.** These methods differ both in the criteria used to select the node for insertion and in the strategy employed to determine its optimal position within the tour. For instance, the Nearest Insertion Heuristic selects the node that minimizes the distance to the current tour and inserts it at the position that minimizes the insertion cost $c_{ikj}$. Given that the Traveling Salesman Problem (TSP) constitutes a particular case of the Vehicle Routing Problem (VRP), numerous constructive heuristics originally developed for the TSP—or adapted from it—can be effectively applied to VRP instances. Notable examples include the Construct-Strike heuristic introduced by Christofides (1973), as well as the Path-Scanning and Augment-Merge methods proposed by Golden et al. (1983).

In contrast to constructive methods, whose objective is to generate a single solution, **metaheuristics that** iteratively refine an initial solution, which may be obtained, for instance, through a constructive heuristic. Each iteration of a local search entails exploring the neighborhood of the current solution in order to identify an improved one. Depending on the strategy employed, this may involve selecting either the first improvement encountered or the best available alternative. The search terminates when no further improvements can be found, in which case the current solution is deemed locally optimal with respect to its neighborhood. Otherwise, the newly identified solution replaces the current one, and the process is repeated. For decades very specific neighboring systems including for example 2-OPT has been proved to be highly efficient to define local search. All of these methods yield approximate solutions of varying quality (depending on the instances) within acceptable computational times. Several attempts have been used to evaluate qualities of different approaches. What emerges from these studies—for instance, (Antosiewicz et al., 2013)—is the ability of these methods to produce optimal or near-optimal solutions within a matter of seconds. (Antosiewicz et al., 2013) report computation times of approximately 100 seconds for instances involving 76 nodes. The computational complexity of these methods has been extensively analyzed in the literature; see, for example, (Punnen et al., 2003). More recently, the work of (Hanif and Ismail, 2017) has extended these findings to larger instances. They note—albeit intuitively—that the quality of solutions produced by approximation methods tends to decline as the instance size increases but for $\simeq 400$ nodes instances they report gap between 2 and 20% to the optimal solution.

In conclusion, it can be stated that for this problem, solutions that are close to the optimal can be obtained using classical methods within reasonable computational times even for large instances.

## 3.3 Warm-start definition

So we are interested in considering one $|\psi_0\rangle$ where a large part of the distribution is concentrated on one quality solution obtained using one classical methods.

**Definition of operator $A$**

Considering one sequence $B = \prod_{k=1}^{n} B_k(\alpha_k)$ where $B_k(\alpha_k) \in \{\cos\alpha_k, \sin\alpha_k\}$,

$$A = \prod_{k=1}^{n} \mathrm{A_k}$$

$$\text{where } \mathrm{A_k} = Id \text{ if } B_k(\alpha_k) = \cos\alpha_k \text{ and } \mathrm{A_k} = Swap_{k,k+1} \text{ if } B_k(\alpha_k) = \sin\alpha_k$$

$$\text{with } k+1 = (k+1)\ mod\ n$$

▯

Let us consider $|\psi_h\rangle = \otimes_{j=1}^{n} \left|\psi_{h_j}\right\rangle$, one initial state fully defined by a high quality solution computed by one non quantum methods.

Let us consider $|\psi_d\rangle = \otimes_{j=1}^{n} \left|\psi_{d_j}\right\rangle$, the specific initial state that creates, after application of the mixer $H_D = \prod_{k=1}^{n} e^{-i.\alpha_k.swap_{k,k+1}}$ , a quantum state where the coefficient of $|\psi_h\rangle$ is exactly defined by $B$.

**Definition of $|\psi_d\rangle = \otimes_{j=1}^{n} \left|\psi_{d_j}\right\rangle$**

$$\left|\psi_{d_j}\right\rangle = (A.|\psi_b\rangle)_j$$

▯

Let us consider one example with $n = 4$ and let us consider:

$$B = \prod_{k=1}^{4} B_k(\alpha_k) = \sin\alpha_4 \cos\alpha_3 \sin\alpha_2 \cos\alpha_1$$

We have

$$A = \prod_{k=1}^{4} \mathrm{A_k} = Swap_{41}.Id.Swap_{23}.Id$$

Let us assume that $|\psi_h\rangle = \left|\psi_{h_1}\psi_{h_2}\psi_{h_3}\psi_{h_4}\right\rangle = |1342\rangle$ is one high quality solution computed using a non quantum approach.

We have

$$A.|\psi_b\rangle = Swap_{41}.Id.Swap_{23}.Id.\left|\psi_{b_1}\psi_{b_2}\psi_{b_3}\psi_{b_4}\right\rangle = \left|\psi_{b_4}\psi_{b_3}\psi_{b_2}\psi_{b_1}\right\rangle$$

Because

$$A.|\psi_b\rangle = |\psi_{b_4}\psi_{b_3}\psi_{b_2}\psi_{b_1}\rangle = |1342\rangle$$

Because

$$|\psi_{d_j}\rangle = (A.|\psi_b\rangle)_j$$

We have $|\psi_{d_4}\rangle = |1\rangle$, $|\psi_{d_3}\rangle = |3\rangle$, $|\psi_{d_2}\rangle = |4\rangle$ and $|\psi_{d_1}\rangle = |2\rangle$

and we conclude that $|\psi_d\rangle = |2431\rangle$.

**Theorem**

$$|\psi_0\rangle = \prod_{k=1}^{n} e^{-i.\alpha_k.swap_{k,k+1}}.|2431\rangle$$

is one quantum state where $\langle\psi_0|1342\rangle = B = \sin\alpha_4 \cos\alpha_3 \sin\alpha_2 \cos\alpha_1$

▯

**Proof.**

$$\prod_{k=1}^{n} e^{-i.\alpha_k.swap_{k,k+1}}.|2431\rangle = \prod_{k=2}^{n} e^{-i.\alpha_k.swap_{k,k+1}} = \cos\alpha_1\ |2431\rangle - i\sin\alpha_1|4231\rangle$$

$$\prod_{k=1}^{n} e^{-i.\alpha_k.swap_{k,k+1}}.|2431\rangle = \prod_{k=3}^{n} e^{-i.\alpha_k.swap_{k,k+1}} = \cos\alpha_2\cos\alpha_1\ |2431\rangle - i\cos\alpha_2\sin\alpha_1|4231\rangle - i.\sin\alpha_2\cos\alpha_1\ |2341\rangle - \sin\alpha_2\sin\alpha_1|4321\rangle$$

$$\prod_{k=1}^{n} e^{-i.\alpha_k.swap_{k,k+1}}.|2431\rangle = \prod_{k=4}^{n} e^{-i.\alpha_k.swap_{k,k+1}} = \cos\alpha_3\cos\alpha_2\cos\alpha_1\ |2431\rangle - i\cos\alpha_3\cos\alpha_2\sin\alpha_1|4231\rangle - i.\cos\alpha_3\sin\alpha_2\cos\alpha_1\ |2341\rangle - \cos\alpha_3\sin\alpha_2\sin\alpha_1|4321\rangle$$

$$-i.\sin\alpha_3\cos\alpha_2\cos\alpha_1\ |2413\rangle - \sin\alpha_3\cos\alpha_2\sin\alpha_1|4213\rangle - \sin\alpha_3\sin\alpha_2\cos\alpha_1\ |2314\rangle + i.\sin\alpha_3\sin\alpha_2\sin\alpha_1|4312\rangle$$

And by consequence

$$\prod_{k=1}^{n} e^{-i.\alpha_k.swap_{k,k+1}}.|2431\rangle = \cos\alpha_4\cos\alpha_3\cos\alpha_2\cos\alpha_1\ |2431\rangle - i\cos\alpha_4\cos\alpha_3\cos\alpha_2\sin\alpha_1|4231\rangle - i.\cos\alpha_4\cos\alpha_3\sin\alpha_2\cos\alpha_1\ |2341\rangle - \cos\alpha_4\cos\alpha_3\sin\alpha_2\sin\alpha_1|4321\rangle - i.\cos\alpha_4\sin\alpha_3\cos\alpha_2\cos\alpha_1\ |2413\rangle - \cos\alpha_4\sin\alpha_3\cos\alpha_2\sin\alpha_1|4213\rangle - \cos\alpha_4\sin\alpha_3\sin\alpha_2\cos\alpha_1\ |2314\rangle + i.\cos\alpha_4\sin\alpha_3\sin\alpha_2\sin\alpha_1|4312\rangle - i\sin\alpha_4\cos\alpha_3\cos\alpha_2\cos\alpha_1\ |1432\rangle - \sin\alpha_4\cos\alpha_3\cos\alpha_2\sin\alpha_1|1234\rangle - \sin\alpha_4\cos\alpha_3\sin\alpha_2\cos\alpha_1\ |1342\rangle + i\sin\alpha_4\cos\alpha_3\sin\alpha_2\sin\alpha_1|1324\rangle - \sin\alpha_4\sin\alpha_3\cos\alpha_2\cos\alpha_1\ |3412\rangle + i\sin\alpha_4\sin\alpha_3\cos\alpha_2\sin\alpha_1|3214\rangle + i\sin\alpha_4\sin\alpha_3\sin\alpha_2\cos\alpha_1\ |4312\rangle + \sin\alpha_4\sin\alpha_3\sin\alpha_2\sin\alpha_1|2314\rangle$$

To conclude $\left\langle 1342 \middle| \prod_{k=1}^{n} e^{-i.\alpha_k.swap_{k,k+1}}.|2431\rangle \right\rangle = -\sin\alpha_4\cos\alpha_3\sin\alpha_2\cos\alpha_1$.

▯

**Parametrization of the mixer $H_D$**

With

$$H_D = \prod_{k=1}^{n} e^{-i.\alpha_k.swap_{k,k+1}}$$

Considering $B = \prod_{k=1}^{n} B_k(\alpha_k)$ we choose $\varepsilon > 0$ and

$$\alpha_k \in \left[0 + \varepsilon; \frac{\pi}{2} - \varepsilon\ \right] \cup \left[\frac{\pi}{2} + \varepsilon\ ; \pi - \varepsilon\right]$$

maximizing $|B|$

The parameter $\varepsilon$ plays a crucial role as it ensures that the wave function is not composed of a single state, which would contradict the objective of exploring the entire set of states representing solutions to the TSP.

□

For example, considering $n = 4$ and $B = -\sin\alpha_4 \cos\alpha_3 \sin\alpha_2 \cos\alpha_1$ , to maximize $|B|$ we can choose $\vec{\alpha}$ such that $\alpha_1 = \alpha_3 \simeq 0$ et $\alpha_2 = \alpha_4 \simeq \frac{\pi}{2}$.

# 4 Warm-Start QAOA

Implementing a QAOA-based approach enables the exploration of the solution space of the Schrödinger equation, whose solutions take the form:

$$|\psi_t\rangle = U(t).\,|\psi_0\rangle$$

$$U(t) = e^{-i.\int_o^t H.du} = e^{-i.t.H} \simeq \prod_{j=1}^{p} e^{-i.\gamma.H_P} e^{-i.\beta.H_D}$$

The target form is of the type $H(\vec{\alpha},\vec{\gamma}) = H_D(\vec{\alpha}) + H_P(\vec{\gamma}\ )$, and depends on the angles $\vec{\alpha}$ and $\vec{\gamma}$ (Fig. 6). These angles will subsequently depend on the iterations performed by the method. In the Hamiltonian thus defined, one must choose the value of the parameter $p$, which represents the number of times the sequence $H_D(\vec{\alpha}) + H_P(\vec{\gamma}\ )$ appears in $H(\vec{\alpha},\vec{\gamma})$. Accordingly, $H(\vec{\alpha},\vec{\gamma})$ depends on $p$ (referred to as the 'depth') and can be written as:

$$e^{-i.H_P(\beta,\gamma)} \simeq e^{-i.H_D(\gamma_p)}.\,e^{-i.H_P(\beta_p)}.\,e^{-i.H_D(\gamma_{p-1})}.\,e^{-i.H_P(\beta_{p-1})} \ldots e^{-i.H_D(\gamma_1)}.\,e^{-i.H_P(\beta_1)}$$

In the Warm-Start QAOA we introduce, the two first step are required to compute $\left|\psi_{d_j}\right\rangle$ and the initial $\vec{\alpha}$ that creates one wave function that concentrate the wave on the solution $|\psi_h\rangle$ leading to $|\psi_0\rangle$.

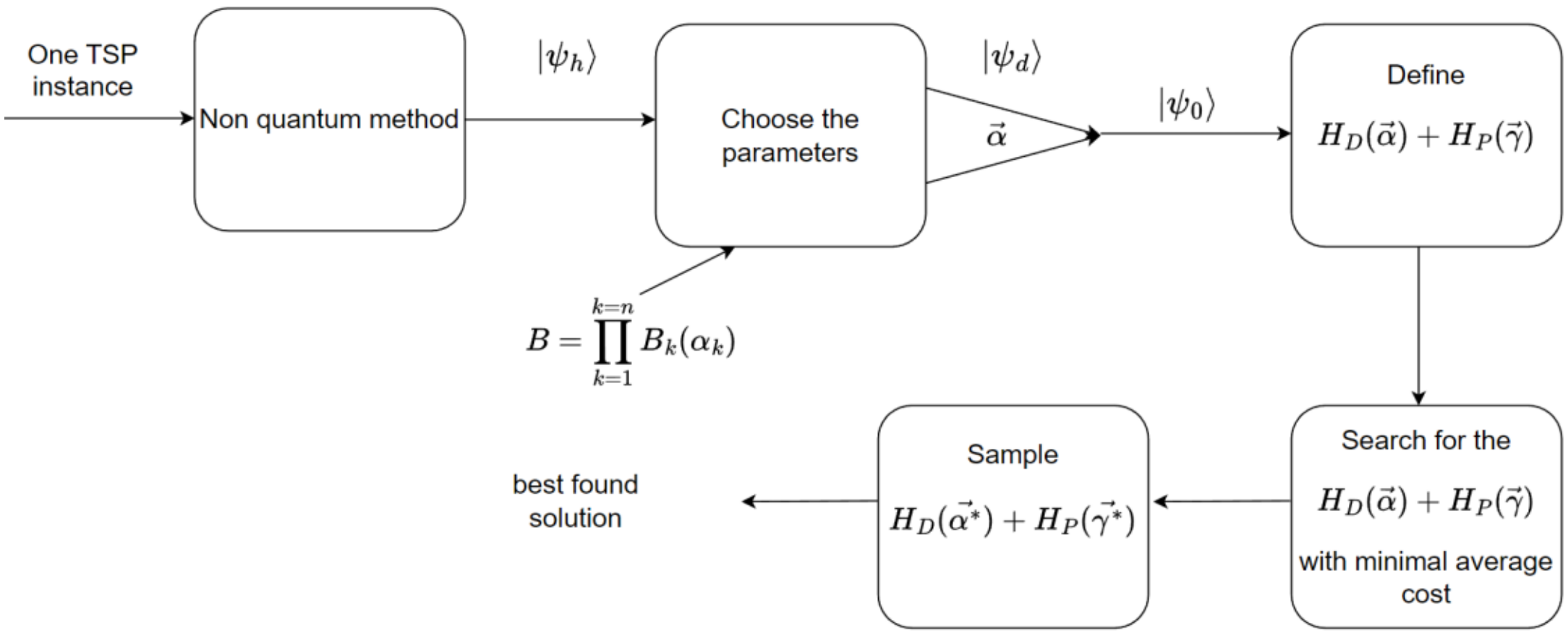

**Fig. 6.** Warm-start QAOA with swap-based mixer

In contrast to QAOA, the vector $\vec{\alpha}$ for each value of $p$ defines $n$ angles, which may differ and are associated with the various swaps in the mixer.

# 5 Numerical experiments with QAOA approach

## 5.1 A 5-customers instance

We consider a Symmetric TSP with 5 nodes and the distance matric below:

$$w = \begin{pmatrix} 0 & 1 & 1.5 & 2 & 7.5 \\ 5 & 0 & 11.5 & 2.5 & 1.75 \\ 0.5 & 6.5 & 0 & 5 & 0.75 \\ 3 & 4.5 & 2 & 0 & 5.25 \\ 1 & 5 & 2.25 & 8 & 0 \end{pmatrix}$$

The quantum modeling requires 21 qubits:

- 20 qubits labelled 0 to 19 to modelize the trip;
- 1 auxiliary qubit (labelled 20).

To obtain a compact representation, we assume that the tour starts at customer 0 and do not represent either the initial or final node. With this convention, the tour 0,1,3,2,4,0 is simply represented by Fig. 7.

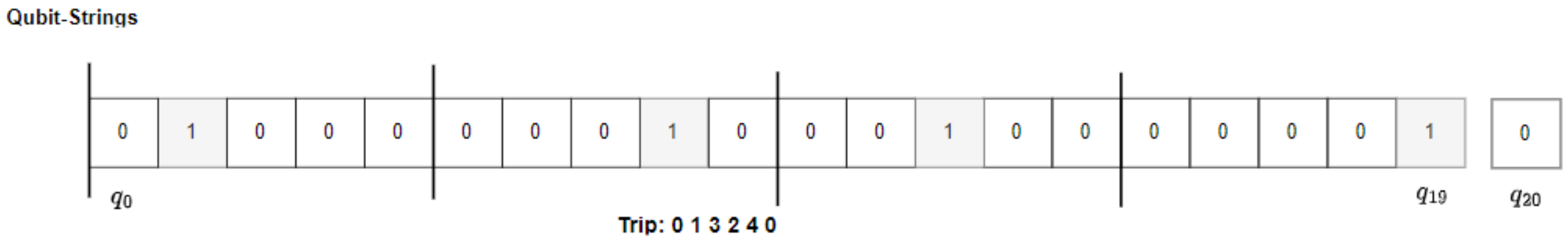

**Fig. 7** Modelization of $0-1-3-2-4-0$

The 5-customers TSP we address corresponds to a problem with 24 solutions, only one of which—the solution 0,1,3,2,4,0—achieves the optimal cost of 7.25. Thus, the 20 qubits used for modeling the solution represent a search space of $2^{20} = 1\,048\,576$ qubit-strings, and among these qubit-strings, the 24 solutions (Table 2) account for only 0.002% of them.

**Table 2.** List of solutions for the 5-customers TSP

| Solutions | | | | | | Cost |
|---|---|---|---|---|---|---|
| 0 | 1 | 2 | 3 | 4 | 0 | 23.75 |
| 0 | 1 | 2 | 4 | 3 | 0 | 24.25 |
| 0 | 1 | 3 | 2 | 4 | 0 | 7.25 |
| 0 | 1 | 3 | 4 | 2 | 0 | 11.50 |
| 0 | 1 | 4 | 2 | 3 | 0 | 13.00 |
| 0 | 1 | 4 | 3 | 2 | 0 | 13.25 |
| 0 | 2 | 1 | 3 | 4 | 0 | 16.75 |
| 0 | 2 | 1 | 4 | 3 | 0 | 20.75 |
| 0 | 2 | 3 | 1 | 4 | 0 | 13.75 |
| 0 | 2 | 3 | 4 | 1 | 0 | 21.75 |
| 0 | 2 | 4 | 1 | 3 | 0 | 12.75 |
| 0 | 2 | 4 | 3 | 1 | 0 | 19.75 |
| 0 | 3 | 1 | 2 | 4 | 0 | 19.75 |
| 0 | 3 | 1 | 4 | 2 | 0 | 11.00 |
| 0 | 3 | 2 | 1 | 4 | 0 | 13.25 |
| 0 | 3 | 2 | 4 | 1 | 0 | 14.75 |
| 0 | 3 | 4 | 1 | 2 | 0 | 24.25 |
| 0 | 3 | 4 | 2 | 1 | 0 | 21.00 |
| 0 | 4 | 1 | 2 | 3 | 0 | 32.00 |
| 0 | 4 | 1 | 3 | 2 | 0 | 17.50 |
| 0 | 4 | 2 | 1 | 3 | 0 | 21.75 |
| 0 | 4 | 2 | 3 | 1 | 0 | 24.25 |
| 0 | 4 | 3 | 1 | 2 | 0 | 32.00 |
| 0 | 4 | 3 | 2 | 1 | 0 | 29.00 |

## 5.2 Python perspective with Qiskit

From an implementation perspective, the Python function named $H$ corresponds to the definition of the initial state and the definition of $H(\vec{\alpha},\vec{\beta})$. $H(\vec{\alpha},\vec{\beta})$ is defined by the procedure def H (beta, gamma, depth) which is composed of one initialization part and of one loop to successively add $H_B$ and $H_P$.

```
def H (beta, gamma, depth):
   qreg_q = QuantumRegister(nb_dequbits, 'q')
   creg_c = ClassicalRegister(nb_dequbits-1, 'c')
   qc = QuantumCircuit(qreg_q, creg_c)
   # depth of QAOA
   p = depth
   # initialization
   for i in range (n-1):
     pos = i*n + (i+1)
     qc.x(qreg_q[pos])

   for i in range (0,p):
     # swap from position i to i+1
     for valeur in range (0,n-2) :
         qc.rx(2*pi/4*beta[i],qreg_q[nb_dequbits-1])
         qcb = CSWAP (beta[i], valeur)
         qc.append(qcb, [nb_dequbits-1] + [i for i in range(0,nb_dequbits-1) ] )
     # swap from the last position to the first one
     qc.rx(2*pi/4*beta[i],qreg_q[nb_dequbits-1])
     qcb = CSWAP2 (beta[i])
     qc.append(qcb, [nb_dequbits-1] + [i for i in range(0,nb_dequbits-1) ] )
```

Part 1

Part 2 : HB definition

```
    # add H_P
    qc_news = H_P (gamma[i])
    qc.append(qc_news, [i for i in range(0,nb_dequbits) ] )

  # measurement
  for i in range(nb_dequbits-1) :
     qc.measure(qreg_q[i], creg_c[i])
  return qc
```

Part 3 : HP definition

The rotation

```
qc.rx(2*pi/4*beta[i],qreg_q[nb_dequbits-1])
```

and

```
qcb = CSWAP (valeur)
```

are the Python implementation of the Algorithm1 to defined $e^{-i.t.U}.|\psi_1,\dots,\psi_{n-1},\psi_n\rangle$ where $U$ is the TSP swap operator. The algorithm 1 takes advantages of one

$CU_0.|\psi_0,\psi_1,\dots,\psi_{n-1},\psi_n\rangle = \cos t\,.|0,\psi_1,\dots,\psi_{n-1},\psi_n\rangle - i.\sin t\,.|1\rangle \otimes U.|\psi_1,\dots,\psi_{n-1},\psi_n\rangle$

that is implemented using the Qiskit method control().

The Python procedure implement iteratively

$$Swap_{ij} = CX_{ij}.CX_{ji}.CX_{ij}$$

for the $n$ qubits.

```
def CSWAP (p):
  qreg_q = QuantumRegister(nb_dequbits-1, 'q')
  qc = QuantumCircuit(qreg_q)

  for i in range (n):
      dep = p*n+i
      arr = p*n+i+n

      qc.cx(qreg_q[dep],qreg_q[arr])
      qc.cx(qreg_q[arr],qreg_q[dep])
      qc.cx(qreg_q[dep],qreg_q[arr])

  c_U = qc.control()
  return c_U
```

The CSWAP() method provides an implementation of the Swap between two customers (Fig. 8.), controlled by an auxiliary qubit.

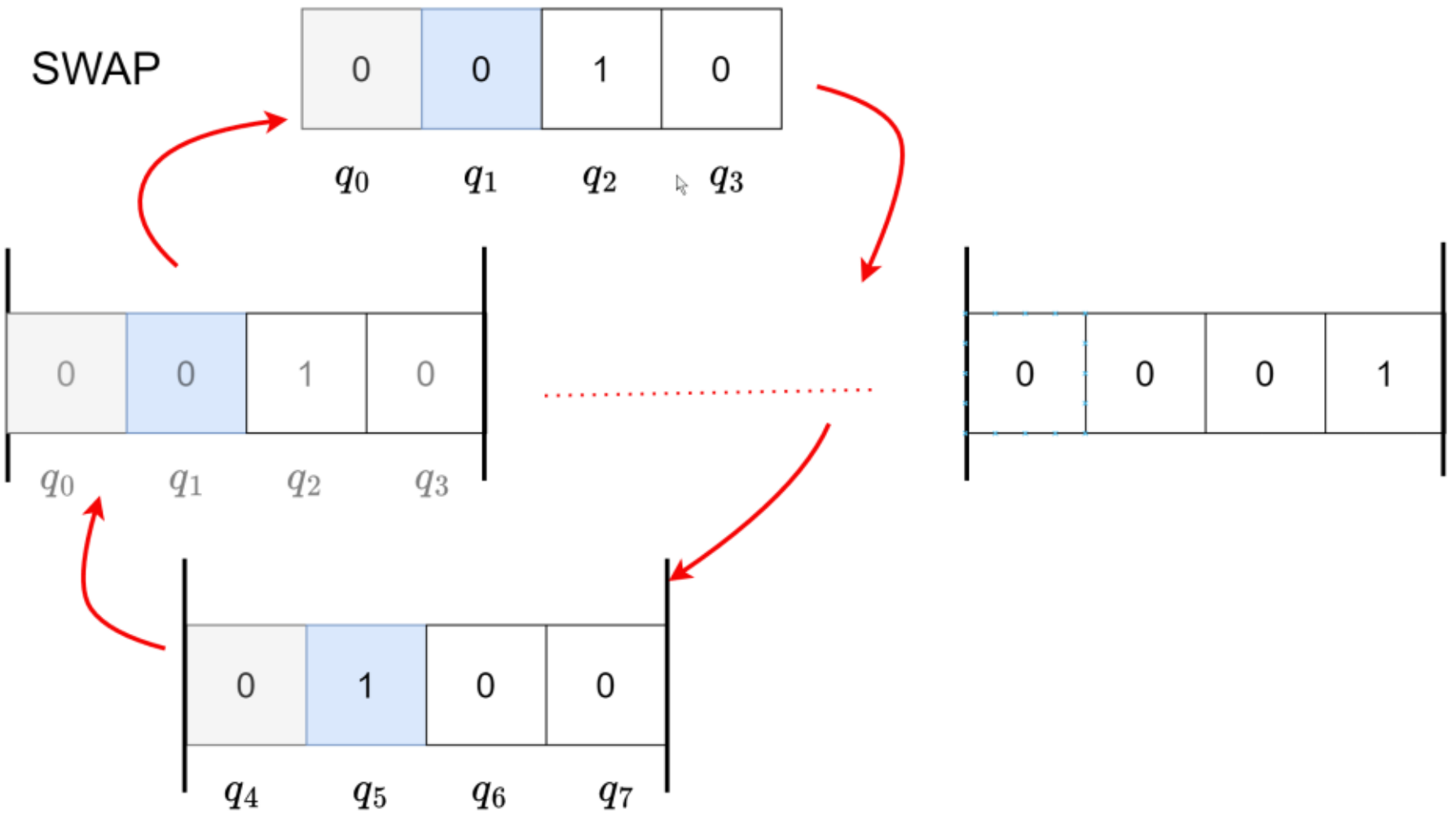

**Fig. 8** Swap between two customers

The `CSWAP2(p)` procedure is similar to `CSWAP()` but defines the swap of the customer in last position with customer in the first position.The function `H_P()` corresponds to the Hamiltonian $H_P$ associated with

$$C(x) = \sum_{i=1}^{n}\sum_{j=1}^{n} w_{ij} \cdot \sum_{p=1}^{n-2} x_{ip}.x_{j,p+1} + \sum_{i=1}^{n} w_{0i}.x_{i,1} + \sum_{i=1}^{n} w_{i,0}.x_{i,n-1}$$

and basically the implementation is introduced below:

```
def H_P (gamma):
   qreg_q = QuantumRegister(nb_dequbits, 'q')
   qc = QuantumCircuit(qreg_q)

   for i in range (1,n) :
     qc.rz(2*gamma*W[0][i],qreg_q[i])

   fin = n*(n-2)+1
   for i in range (n-1):
       qc.rz(2*gamma*W[i][0],qreg_q[i+fin])

   for i in range (1,n-1):
       for j in range (1,n):
           if (j!=i) :
               for p in range (0,n-2):
                   dep = p*(n) + i
                   arr = (p+1)*(n) + j

                   qc.cx(qreg_q[dep],qreg_q[arr])
                   qc.rz(2*gamma*1/4*W[i][j],qreg_q[arr])
                   qc.cx(qreg_q[dep],qreg_q[arr])

                   qc.rz(-2*gamma*1/4*W[i][j],qreg_q[dep])
                   qc.rz(-2*gamma*1/4*W[i][j],qreg_q[arr])

   return qc
```

The calls to the function *H* are encapsulated in the function *f*, which is generated by the procedure `definir_la_fonction_a_minimiser(p)`.

```
def definir_la_fonction_a_minimiser(p):
    def f (theta):
        print(theta)
        beta  = theta[:p]
        gamma = theta[p:]
        qc = H (beta, gamma, p)
        qc_compiled = transpile(qc, backend)
        job_sim = backend.run(qc_compiled, shots=NUM_SHOTS)
        result_sim = job_sim.result()
        counts = result_sim.get_counts(qc_compiled)
        les_resultats = inversion_affichage(counts)
        res = evaluer_H(les_resultats)
        return res
    return f
```

The optimization process simply begins with a call to `minimize(…)` using the ad-hoc parameters.

```
    p = 2 # choose the QAOA depth
    la_fonction_objectif = definir_la_fonction_a_minimiser(p)
    init_point = np.array([1, 1, 1, 1])
    res = la_fonction_objectif (init_point)
```

```python
    res_sample = minimize(la_fonction_objectif,
                          init_point,
                          method='Cobyla', # Nelder-Mead Cobyla BFGS
                          options={'maxiter':250, 'disp': True}
                          )
    print(res_sample)
```

## 5.2 Experiments

The tests were carried out using the Cobyla method with 250 iterations, and the results are shown in Table 3. The solution with a cost of 7.25 accounts for 15.1% of the distribution, whereas a uniform random selection among the 24 solutions would have yielded a probability of 4.16%. The 250 iterations of Cobyla achieved an amplification factor of approximately 4.

**Table 3.** Probabilities after optimization

| Cost | Probability |
|---|---|
| 7.25 | 15.1 |
| 11 | 5.3 |
| 11.5 | 8.4 |
| 12.75 | 2.5 |
| 13 | 8.7 |
| 13.25 | 3.1 |
| 13.75 | 2.9 |
| 14.75 | 4.2 |
| 16.75 | 2.7 |
| 17.5 | 1.4 |
| 19.75 | 3.9 |
| 20.75 | 6 |
| 21 | 4.6 |
| 21.75 | 3.7 |
| 23.75 | 7 |
| 24.25 | 6.7 |
| 29 | 6.5 |
| 32 | 7.3 |

Thanks to the swap based mixer we have introduced all the distribution is concentrated on the qubit-strings that modelized only the 24 solutions of the problem as stressed on Fig. 9.

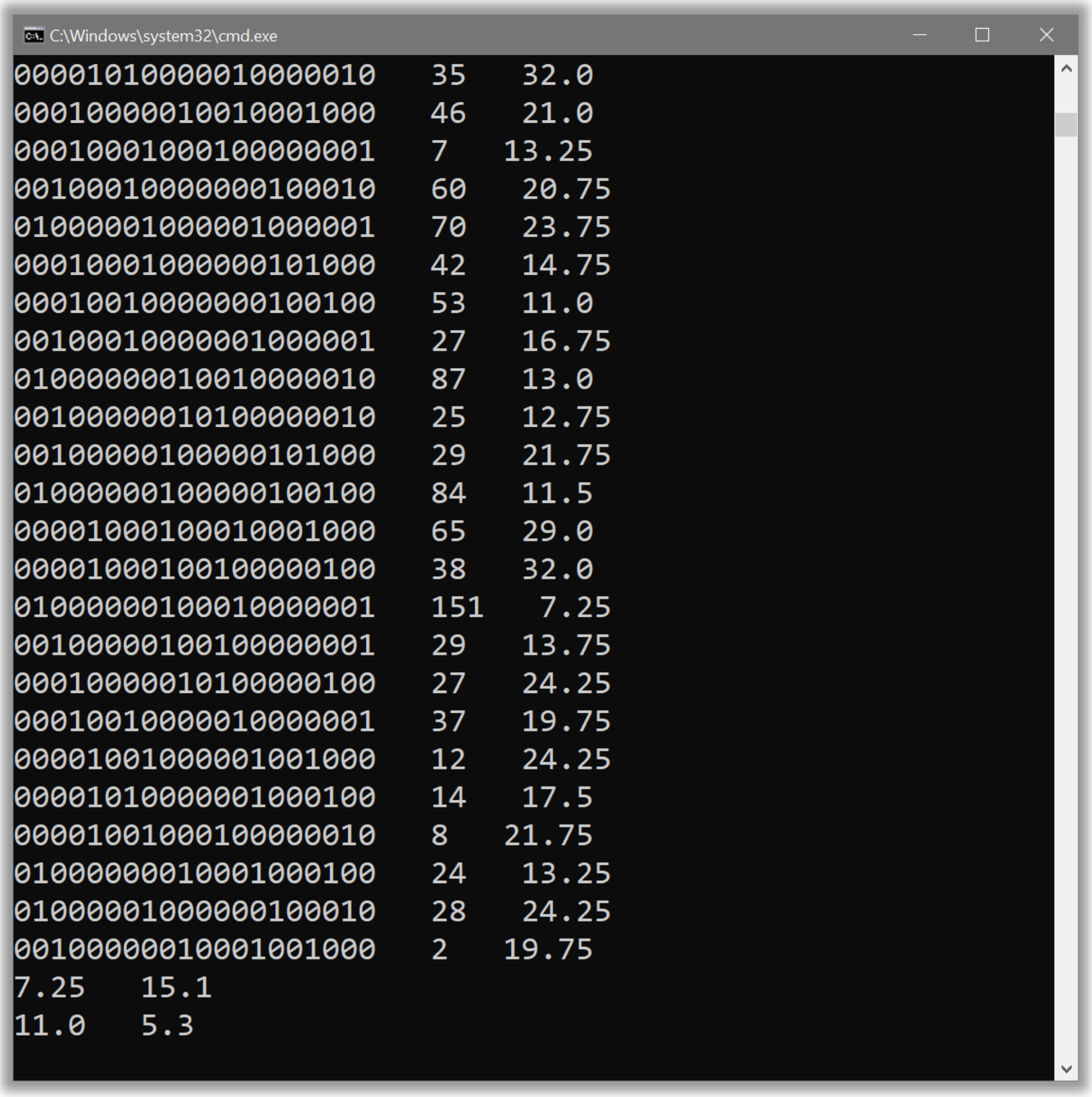

**Fig. 9.** The distribution of probability concentrated on solutions

# 6 Numerical experiments with warm-start QAOA approach

As with all problem instances, the overall performance of QAOA depends not only on the circuit depth $p$, but also on the optimization method and the cost function used—an observation already highlighted for the SAT problem, for instance, in (Fleury and Lacomme, 2023). In the following, we focus on the ability to initialize QAOA from a known quality solution. The choice of both the optimization methods and the cost functions has not been subject to specific tuning, but rather based on straightforward, heuristic selections. All the experiments have been carried out with $N = 200$ shots.

## 6.2 Experiments 1 with $p = 1$

Let us assume that one heuristic provides the solution $|\psi_h\rangle = |1342\rangle$ and let us consider a very specific function $B = \prod_{k=1}^{n} B_k(\alpha_k) = \sin\alpha_4 \cos\alpha_3 \sin\alpha_2 \cos\alpha_1$. Considering the definition, we have $|\psi_d\rangle = 2431$ leading to:

$$
\begin{aligned}
|\psi_0\rangle = {} & \cos\alpha_4 \cos\alpha_3 \cos\alpha_2 \cos\alpha_1\ |2431\rangle - i\cos\alpha_4 \cos\alpha_3 \cos\alpha_2 \sin\alpha_1 |4231\rangle \\
& - i.\cos\alpha_4 \cos\alpha_3 \sin\alpha_2 \cos\alpha_1\ |2341\rangle - \cos\alpha_4 \cos\alpha_3 \sin\alpha_2 \sin\alpha_1 |4321\rangle \\
& - i.\cos\alpha_4 \sin\alpha_3 \cos\alpha_2 \cos\alpha_1\ |2413\rangle - \cos\alpha_4 \sin\alpha_3 \cos\alpha_2 \sin\alpha_1 |4213\rangle \\
& - \cos\alpha_4 \sin\alpha_3 \sin\alpha_2 \cos\alpha_1\ |2314\rangle + i.\cos\alpha_4 \sin\alpha_3 \sin\alpha_2 \sin\alpha_1 |4312\rangle \\
& - i\sin\alpha_4 \cos\alpha_3 \cos\alpha_2 \cos\alpha_1\ |1432\rangle - \sin\alpha_4 \cos\alpha_3 \cos\alpha_2 \sin\alpha_1 |1234\rangle \\
& - \sin\alpha_4 \cos\alpha_3 \sin\alpha_2 \cos\alpha_1\ |1342\rangle \\
& + i\sin\alpha_4 \cos\alpha_3 \sin\alpha_2 \sin\alpha_1 |1324\rangle - \sin\alpha_4 \sin\alpha_3 \cos\alpha_2 \cos\alpha_1\ |3412\rangle \\
& + i\sin\alpha_4 \sin\alpha_3 \cos\alpha_2 \sin\alpha_1 |3214\rangle + i\sin\alpha_4 \sin\alpha_3 \sin\alpha_2 \cos\alpha_1\ |4312\rangle \\
& + \sin\alpha_4 \sin\alpha_3 \sin\alpha_2 \sin\alpha_1 |2314\rangle
\end{aligned}
$$

To maximize $B$ we choose $\alpha_3 = \alpha_1 = \epsilon$ and $\alpha_2 = \alpha_4 = \frac{\pi}{2} - \epsilon$ where $\epsilon$ is one parameter used to control the concentration of the wave function on $|1342\rangle$. When $\epsilon = 0$ the wavefunction is composed of solution cost 11.5 that is the cost of $|1342\rangle$. By increasing the value of $\epsilon$, the wave function is progressively altered, leading to more uniform distributions, as shown in Table 4.

**Table 4.** Wavefunction depending on $\epsilon$

| | $\epsilon = 0$ | | $\epsilon = 0.1$ | | $\epsilon = 0.2$ | | $\epsilon = 0.3$ | |
|---|---|---|---|---|---|---|---|---|
| Cost | Nb of shoots | Percentage | Nb of shoots | Percentage | Nb of shoots | Percentage | Nb of shoots | Percentage |
| 7.25 | 0 | 0 | 0 | 0 | 0 | 0 | 0 | 0 |
| 11 | 0 | 0 | 0 | 0 | 0 | 0 | 0 | 0 |
| **11.5** | **200** | **100** | **174** | **87** | **127** | **63.5** | **56** | **28** |
| 12.75 | 0 | 0 | 10 | 5 | 16 | 8 | 29 | 14.5 |
| 13 | 0 | 0 | 0 | 0 | 0 | 0 | 0 | 0 |
| 13.25 | 0 | 0 | 0 | 0 | 1 | 0.5 | 13 | 6.5 |
| 13.75 | 0 | 0 | 2 | 1 | 23 | 11.5 | 33 | 16.5 |
| 14.75 | 0 | 0 | 0 | 0 | 0 | 0 | 0 | 0 |
| 16.75 | 0 | 0 | 0 | 0 | 0 | 0 | 0 | 0 |
| 17.5 | 0 | 0 | 0 | 0 | 0 | 0 | 0 | 0 |
| 19.75 | 0 | 0 | 0 | 0 | 2 | 1 | 14 | 7 |
| 20.75 | 0 | 0 | 0 | 0 | 0 | 0 | 0 | 0 |
| 21 | 0 | 0 | 0 | 0 | 0 | 0 | 0 | 0 |
| 21.75 | 0 | 0 | 12 | 6 | 22 | 11 | 28 | 14 |
| 23.75 | 0 | 0 | 0 | 0 | 1 | 0.5 | 5 | 2.5 |
| 24.25 | 0 | 0 | 1 | 0.5 | 3 | 1.5 | 11 | 5.5 |
| 29 | 0 | 0 | 0 | 0 | 0 | 0 | 4 | 2 |
| 32 | 0 | 0 | 1 | 0.5 | 5 | 2.5 | 7 | 3.5 |

The experiments are carried out with the Cobyla solver using 25 iterations only with $p = 1$ for minimization of the average cost and as illustrated in Figure 10, the COBYLA method exhibits rapid convergence towards an average cost of 7.9.

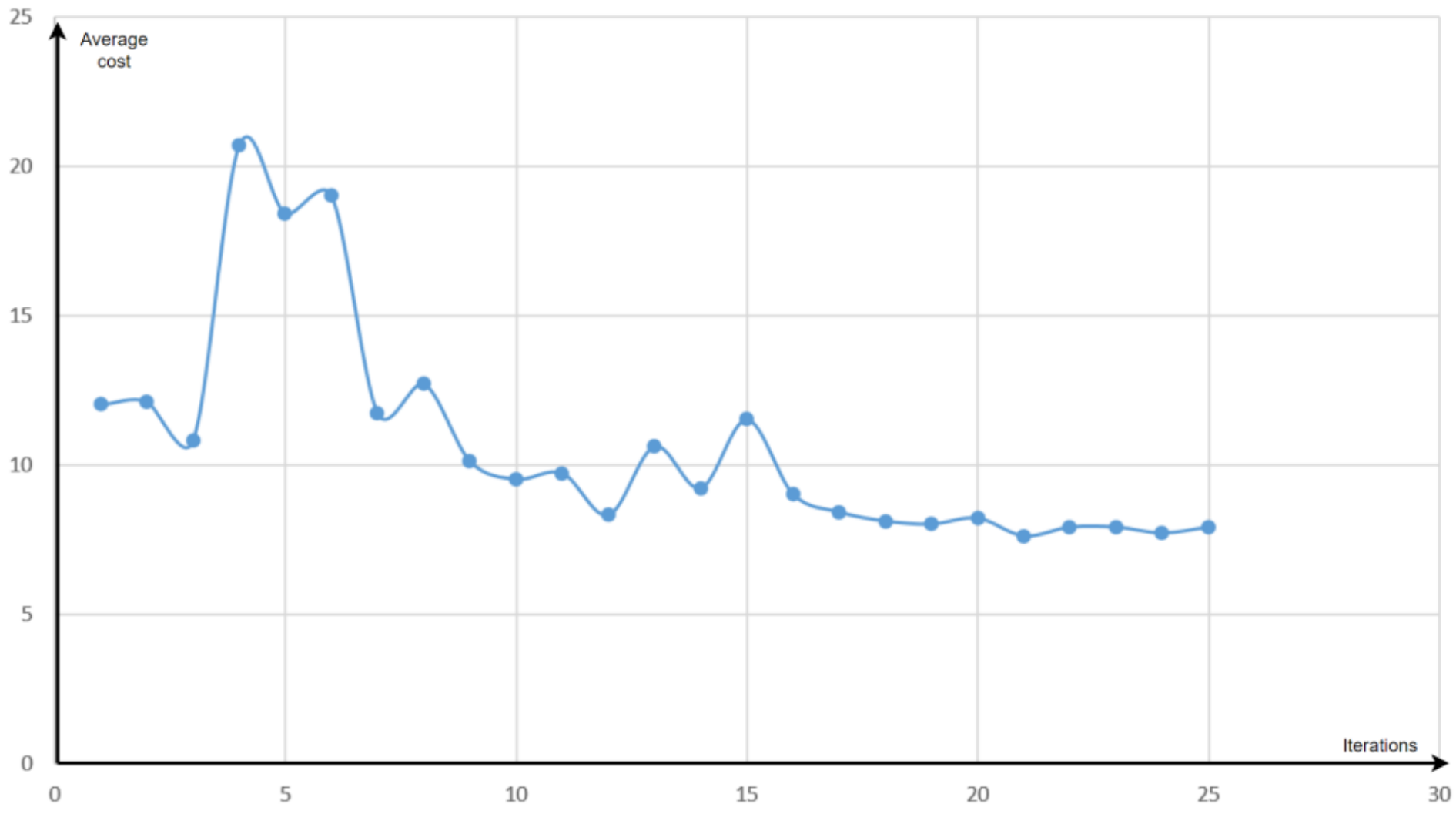

**Fig. 10.** Visualization of the average cost over iterations with 25 iterations of Cobyla

**Table 5.** Probabilities after optimization (25 iterations)

| Cost | Probabilities |
|---|---|
| 7.25 | 94.0 |
| 11 | 0.0 |
| 11.5 | 0.0 |
| 12.75 | 3.5 |
| 13 | 0.0 |
| 13.25 | 0.0 |
| 13.75 | 0.0 |
| 14.75 | 0.0 |
| 16.75 | 0.0 |
| 17.5 | 0.0 |
| 19.75 | 0.0 |
| 20.75 | 0.0 |
| 21 | 0.0 |
| 21.75 | 1.5 |
| 23.75 | 0.0 |
| 24.25 | 0.0 |
| 29 | 0.0 |
| 32 | 1.0 |

The results of Table 5 and Fig. 11. prove that about 94% of the distribution is now concentrated on the optimal solution that value 7.25.

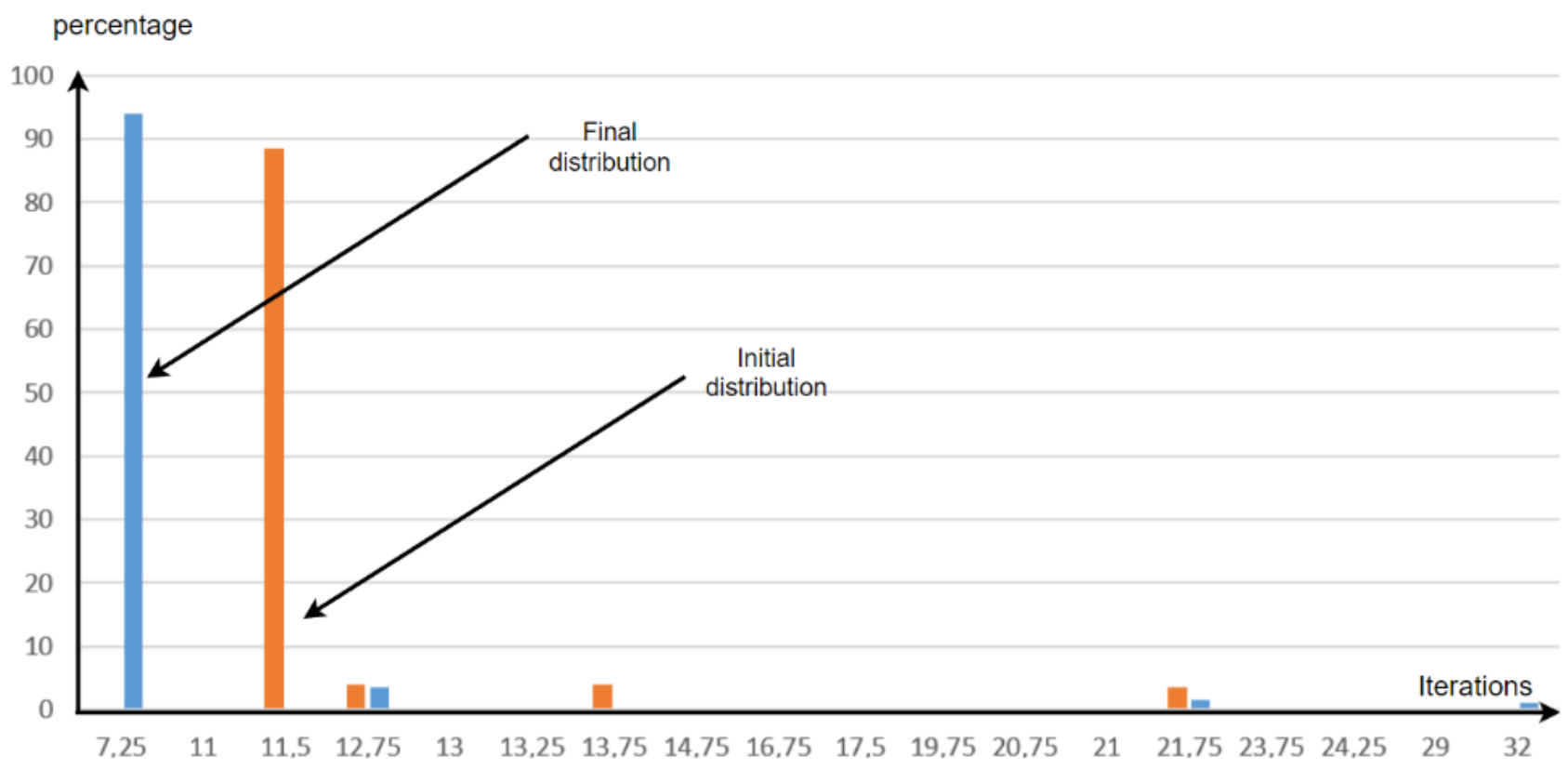

**Fig. 11.** Visualization of the distribution of probability over cost avec 25 iterations of Cobyla

## 6.2 Experiment 2 with $p = 3$

Let us assume that one heuristic provides the solution $|\psi_h\rangle = |1342\rangle$ and let us consider a very specific function $B = \prod_{k=1}^{n} B_k(\alpha_k) = \sin\alpha_4 \sin\alpha_3 \sin\alpha_2 \cos\alpha_1$. Considering the definition, we have $|\psi_d\rangle = |2134\rangle$ leading to:

$$|\psi_0\rangle = \cos\alpha_4 \cos\alpha_3 \cos\alpha_2 \cos\alpha_1 \ |2134\rangle - i\cos\alpha_4 \cos\alpha_3 \cos\alpha_2 \sin\alpha_1 \,|1234\rangle - i.\cos\alpha_4 \cos\alpha_3 \sin\alpha_2 \cos\alpha_1 \ |2314\rangle - \cos\alpha_4 \cos\alpha_3 \sin\alpha_2 \sin\alpha_1 \,|1324\rangle$$

$$-i.\cos\alpha_4 \sin\alpha_3 \cos\alpha_2 \cos\alpha_1 \ |2143\rangle - \cos\alpha_4 \sin\alpha_3 \cos\alpha_2 \sin\alpha_1 \,|1243\rangle - \cos\alpha_4 \sin\alpha_3 \sin\alpha_2 \cos\alpha_1 \ |2341\rangle + i.\cos\alpha_4 \sin\alpha_3 \sin\alpha_2 \sin\alpha_1 \,|1342\rangle$$

$$-i\sin\alpha_4\cos\alpha_3\cos\alpha_2\cos\alpha_1\ |4132\rangle - \sin\alpha_4\cos\alpha_3\cos\alpha_2\sin\alpha_1\,|4231\rangle$$
$$-\sin\alpha_4\cos\alpha_3\sin\alpha_2\cos\alpha_1\ |4312\rangle + i\sin\alpha_4\cos\alpha_3\sin\alpha_2\sin\alpha_1\,|4321\rangle$$

$$-\sin\alpha_4\sin\alpha_3\cos\alpha_2\cos\alpha_1\ |3142\rangle + i\sin\alpha_4\sin\alpha_3\cos\alpha_2\sin\alpha_1\,|3241\rangle$$
$$+ i\sin\alpha_4\sin\alpha_3\sin\alpha_2\cos\alpha_1\ |1342\rangle + \sin\alpha_4\sin\alpha_3\sin\alpha_2\sin\alpha_1\,|2341\rangle$$

By consequence

$$\langle\psi_0|1342\rangle = i.\cos\alpha_4\sin\alpha_3\sin\alpha_2\sin\alpha_1 + i\sin\alpha_4\sin\alpha_3\sin\alpha_2\cos\alpha_1$$

To obtain a great part of the probability concentrated on $|1234\rangle$, it is possible to parametrized $B$ properly defining for example by

$$\alpha_4 = \alpha_3 = \alpha_2 \simeq \frac{\pi}{2} \pm \epsilon$$

$$\alpha_1 \simeq \pm\epsilon$$

**Table 6.** Probabilities before and after optimization (250 iterations)

| Cost | Initial probabilities | Final Probabilities |
|---:|---:|---:|
| 7.25 | 0.0 | 60.6 |
| 11 | 0.0 | 0.6 |
| 11.5 | 78.0 | 0.3 |
| 12.75 | 0.0 | 0.7 |
| 13 | 10.0 | 0.8 |
| 13.25 | 0.0 | 18.7 |
| 13.75 | 0.0 | 0.7 |
| 14.75 | 0.0 | 0.1 |
| 16.75 | 0.0 | 0.5 |
| 17.5 | 0.0 | 4.8 |
| 19.75 | 0.0 | 5.1 |
| 20.75 | 0.0 | 0.1 |
| 21 | 0.0 | 0.7 |
| 21.75 | 0.0 | 0.1 |
| 23.75 | 12.0 | 0.1 |
| 24.25 | 0.0 | 2.7 |
| 29 | 0.0 | 1.4 |
| 32 | 0.0 | 1.9 |

With $\epsilon = 0.3$ for $p = 1$ and the values for $p = 2,3$ are chosen to define one uniform distribution $\alpha_4 = \alpha_3 = \alpha_2 = \alpha_1 = 0$ to use the identity operator of the exponential.

The objective function has been defined by:

$$f = K \times Min + (N - n_0)$$

where
$K$ is a large integer number ($K = 100000$)
$Min$ is the lower value found over the $N$ shoots
$n_0$ is the total number of shoots on $Min$

The results introduced in table 6 proves that the method converge to one distribution of probabilities concentrated on 7.25.

# 7 Conclusion

Following the research trend of Hadfield (2018). this paper introduces a swap-based mixer specifically designed for the TSP. enabling the exploration of qubit strings that represent TSP solutions exclusively. We present numerical experiments demonstrating that the experimental results align with theoretical expectations. The TSP serves as the cornerstone of a broad class of routing problems in operations research. and numerous neighboring systems. such as 2-OPT and 3-OPT. have been introduced over the decades. proving to be highly effective in local search strategies. Such operators could offer new avenues for defining mixers

**Remark.** Authors used ChatGPT to improve English.